\documentclass[bibyear]{aa} 

\usepackage{graphicx}
\usepackage{txfonts}
\usepackage{hyperref}
\usepackage{multirow}
\usepackage{lscape}
\usepackage{amsmath}
\usepackage{placeins}
\usepackage{xcolor}
\usepackage{natbib}

\begin{document}

   \title{ Spiral shocks induced in galactic gaseous disk: hydrodynamic understanding of observational properties of spiral galaxies}


    \author{Ramiz Aktar\inst{\ref{XMU},\ref{NTHU}} 
                   \and Li Xue\inst{\ref{XMU}}
                   \and Li-Xin Zhang\inst{\ref{XMU}}
                   \and Jing-Yi Luo\inst{\ref{XMU}}
          }
    
   \institute{Department of Astronomy, Xiamen University, Xiamen, Fujian 361005, People’s Republic of China\\ \label{XMU} 
   \and Department of Physics and Institute of Astronomy, National Tsing Hua University, 30013 Hsinchu, Taiwan \\ \email{ramizaktar@gmail.com} \label{NTHU}
}

\titlerunning{Spiral shocks}
\authorrunning{Aktar et al.}


 
  \abstract
   {We investigate the properties of spiral shocks in a steady, adiabatic, non-axisymmetric, self-gravitating, mass-outflowing accretion disk around a compact object. }
   { We obtain the accretion-ejection solutions in a galactic disk and apply them to the spiral galaxies to investigate the possible physical connections between some galaxy observational quantities.}
   {The self-gravitating disk potential is considered following \citet{Mestel-63} prescription. The spiral shock induced accretion-ejection solutions are obtained following the point-wise self-similar approach \citep{Aktar-etal21}.}
   {We observe that the self-gravitating disk profoundly affects the dynamics of the spiral structure of the disk and the properties of the spiral shocks. We find that the observational dispersion between the pitch angle and shear rate and between the pitch angle and star formation rate in spiral galaxies contains some important physical information. }
   {There are large differences of star formation rates among galaxies with similar pitch angle, may be explained by the different star formation efficiencies caused by the distinct galactic ambient conditions.}

   \keywords{Galaxy: disk --
                Galaxies: spiral --
                Galaxies: star formation -- Shock waves
               }

   \maketitle
%

\section{Introduction} 

The spiral structure is a long-term and fascinating topic in the observational and theoretical study of accretion disks. In the observation, there are many pieces of evidence for the existence of the spiral structure in accretion disks \citep{Steeghs-etal97, Neustroev-Borisov98, Pala-etal19, Baptista-etal20, Lee-etal20}. It has become a general consensus that the spiral shock wave induces this spiral structure in the accretion disk. However, the origin of the shock may correspond to many different mechanisms. In theory, \citet{Michel-84} firstly proposed the spiral shock in accretion disks as an effective angular momentum transfer mechanism. \citet{Sawada-etal86a, Sawada-etal86b} performed two-dimensional hydrodynamic simulations of the Roche lobe overflow in a semi-detached binary system to confirm the formation of the spiral shock and the angular momentum transfer in an accretion disk. From the new millennium onwards, with the progress of enormous computational facilities, more and more three-dimensional simulations, which include the spiral shock, have been investigated for the accretion in a binary system in many different studies \citep{Makita-etal00, Molteni-etal01, Ju-etal16, Ju-etal17, Xue-etal21}.

Though the solution of numerical simulation is closer to the physical reality, the insight of a simplified model of intrinsic physical laws still plays an essential role in developing a theory. In the theoretical field on the spiral shock of accretion disks, \citet{Spruit-87} first introduced the radial self-similar simplification for the steady accretion flow in an inertial frame. The same simplification has also been adopted in subsequent theoretical studies of accretion disks (e.g., \citet{Chakrabarti-90a, Narayan-Yi94}). It is worth mentioning that it is a common feature that the Newtonian gravitational potential has been used in these studies, which maintains the mathematical self-consistency of self-similar solutions at different radii. In addition, \citet{Narayan-Yi94} pointed out that this kind of radial self-similar solutions under the Newtonian potential are piece-wise valid, which can only match the simulations in the middle radial region of accretion disks where there is less effect from the inner and outer boundaries. Though, they can be applied to all available radii mathematically. Following these theoretical studies, we extended the spiral shock model presented by \citet{Spruit-87} and further improved by \citet{Chakrabarti-90a} from the single star in an inertial frame to the binary system in a non-inertial corotating frame as well as involved the mass outflow induced by spiral shocks \citep{Aktar-etal21}. Accordingly, the Newtonian potential has been replaced by the Roche potential as well as the Coriolis force. This allows us to involve the effects of the binary system on the spiral shock in our model, but our self-similar solution degenerates to become point-wise valid because it is no longer to keep the separation of variables valid at different radii. 

On the other hand, the existence of shock waves in an axisymmetric accretion flow and their implication has been extensively studied in literature both analytically and numerically \citep{Fukue-87, Chakrabarti89, Lu-etal99, Becker-Kazanas01, Fukumura-Tsuruta04, Chakrabarti-Das04, Sarkar-Das16, Sarkar-etal18, Dihingia-etal18, Dihingia-etal19a, Dihingia-etal19b, Sarkar-etal20}. Due to the shock transition, the post-shock matter becomes very dense and hot (known as post-shock corona (PSC), see \citet{Aktar-etal15}). As a result, a part of the accreting matter is ejected as mass outflow from the disk due to the excess thermal gradient force across the shock. The accretion-ejection process has been widely investigated based on the shock compression model considering an axisymmetric accretion flow assumption \citep{Chattopadhyay-Das07, Das-Chattopadhyay08, Kumar-Chattopadhyay13, Aktar-etal15, Aktar-etal17, Aktar-etal19}. In the same spirit, \citet{Aktar-etal21} investigated mass outflow from the disk induced by spiral shock compression in a non-axisymmetric accretion flow. 

In another astrophysical field, the spiral structure in galaxies has also been investigated for a long time. The number of spiral arms and pitch angle (PA, the angle between the tangent and azimuthal directions on the spiral arm) are both essential criteria of Hubble's scheme for classifying galaxies \citep{Hubble26}. \citet{Lin-Shu64} proposed the famous density wave theory to explain the formation and preservation of spiral arms in galaxies. \citet{Woodward-76} performed a two-dimensional hydrodynamical simulation to demonstrate the mechanism of star formation (SF) in the density wave theory. \citet{Elmegreen-79} proposed that the interstellar matter flows through the spiral density wave, becomes shocked, and then collapses by its self-gravity. \citet{Block-etal97} studied the spiral arms of M51 and found evidences that the gravitational collapse of the shocked gas triggers the SF in spiral arms.

Inspired by these studies, based on our self-similar model of spiral shocks \citep{Aktar-etal21}, we are encouraged to investigate the possible correlation between the star formation rate (SFR) and the characteristic quantity of spiral arms, PA, in spiral galaxies. Since the galactic gaseous disk is self-gravitating, our model must be modified to adapt to this new situation (see Section \ref{model-description}) and involve the SF as a special kind of mass outflow from the gaseous disk (see Section \ref{application}). Additionally, since the self-gravity of disk depends on the specific disk mass distribution, the self-similar solution of our model would be locally point-wise (radius-wise) valid, which enables us to apply some similar methodologies from our previous work \citep{Aktar-etal21}.

We organize the paper as follows. In section \ref{model-description}, we present the description of the model and governing equations. In section \ref{results}, we discuss the results of our model in detail. In section \ref{application}, we apply our model to understand the dispersion between galactic observational quantities. Finally, we draw the concluding remarks in section \ref{conclusion}.

\section{Model Description}
\label{model-description}

We consider a steady, adiabatic, non-axisymmetric accretion flow around a compact star. Here, we assume that the effect of gravity on the accretion disk is significant enough compared to the central object. Therefore, we consider the self-gravitating disk in this paper. We also adopt the spiral shock model proposed by \citet{Chakrabarti-90b}. In this work, we simultaneously solve the radial and the azimuthal components of momentum equations and consider that the accretion flow is in vertical hydrostatic equilibrium throughout the disk.

\subsection{Governing Equations}

In this paper, we write the governing equations in cylindrical coordinates on the equatorial plane. The governing equations are as follows\\
(i) The radial momentum conservation equation:
\begin{equation}\label{radial_eq}
v_r\frac{\partial v_r}{\partial r} + \frac{v_{\phi}}{r} \frac{\partial v_r}{\partial \phi} +\frac{1}{\rho} \frac{\partial P}{\partial r}  - \frac{v_{\phi}^2}{r} + \frac{\partial \Phi}{\partial r} =0,
\end{equation}
(ii) The azimuthal momentum equation:
\begin{equation}\label{azimuthal_eq}
v_r \frac{\partial v_\phi}{\partial r} + \frac{v_\phi}{r} \frac{\partial v_\phi}{\partial \phi} + \frac{v_\phi v_r}{r} + \frac{1}{r \rho} \frac{\partial P}{\partial \phi}  =0,
\end{equation}
(iii) The continuity equation:
\begin{equation} \label{continuity_eq}
\frac{\partial }{\partial r}(hv_r \rho r) + \frac{\partial }{\partial \phi}(h \rho v_{\phi}) =0, 
\end{equation}
and finally\\
(iv) The vertical pressure balance equation:
\begin{equation}\label{vertical_eq}
\frac{1}{\rho} \frac{\partial P}{\partial z} =\left( \frac{\partial \Phi}{\partial z}\right)_{z<<r},
\end{equation}
where $r$, $\phi$, $v_r$, $v_\phi$, $P$, $\rho$, and $2h$ are the radial coordinate, azimuthal coordinate, the radial component of velocity, the azimuthal component of velocity, gas pressure, the density of the flow, and local vertical thickness, respectively. The $\Phi$ in equation (\ref{radial_eq} and \ref{vertical_eq}) is the total gravitational potential due to the compact object present at the center of the disk and self-gravitational potential due to the disk material. The expression of $\Phi$ is given in section \ref{self-gravity-sec}. We also use the adiabatic equation of state $P=K\rho^{\gamma}$, where $K$ is the measure of the entropy of the flow. $\gamma =1+ \frac{1}{n}$ is the adiabatic index, and $n$ represents polytropic index of the flow.

\subsection{Self-gravitating disk}
\label{self-gravity-sec}

In an accretion disk, the gravitational field is generally dominated by the central compact object, but in some cases, the disk's self-gravity can also produce a significant effect. The contribution of gravitational field due to disk depends on the matter distribution through Poisson's equation. For an infinitesimally thin disk, the relation between the surface density of the disk ($\Sigma(r)$) and the disk gravitational field $(\Phi_d)$ can be written using the complete elliptic integrals of the first kind \citep{Lodato-07}. The integral form is quite complicated to handle analytically. However, there is a particular simplified relation between $\Sigma$ and $\Phi_d$ at the disk midplane proposed by \citet{Mestel-63}. In this work, we consider the gravitational force due to self-gravitating disk, and is given by
$$
\frac{\partial \Phi_d}{\partial r} = 2\pi G \Sigma(r)
\eqno(5)
$$
where, $\Sigma = 2 \rho h$ is the surface density of the disk. Therefore, the total gravitational force in the presence of a self-gravitating disk, as well as the central compact object, is given by
\begin{align*} \label{potential_eqn}
\frac{\partial \Phi}{\partial r}  = \frac{\partial}{\partial r} (\Phi_c + \sigma \Phi_d)  = \frac{GM}{r^2} + \sigma~2\pi G \Sigma(r)
\tag{6}
\end{align*}
where $\Phi_c$ and $\Phi_d$ are the gravitational potentials due to the compact object at the center of the disk and due to the self-gravity of the disk, respectively. Here, $G$ is the Gravitational constant. We also introduce a constant factor $\sigma$. For which, $\sigma = 0$ implies non self-gravitating disk \citep{Chakrabarti-90b}, and $\sigma=1$ introduces the effect of self-gravity. In this paper, we use the unit system $G=M=c=1$ throughout; otherwise, it is stated.

It is to be emphasized that in reality, the gravity torque generated by the spiral arm of the spiral galaxy is inevitable \citep{Block-etal-02, Block-etal-04, Tiret-Combes-08}. However, in the present work, we ignore the effect of gravity torque in the presence of the spiral arm in equation \ref{potential_eqn}. Further, it is to be mentioned that the spiral galactic disk is composed of visible and invisible matter such as gaseous matter, stars, dark matter, etc. In our present theoretical model, we assume the galactic disk is predominately dominated by gaseous matter, and calculation is independent of mass. However, we consider the total mass visible or invisible within the radius $r$ when we derive the observational data from the circular velocity curve of spiral galaxies (see section \ref{application}). Therefore, our calculation considers the gravity contribution from the stellar component and invisible mass implicitly.

\subsection{Flow equations in spiral coordinates using self-similar conditions}

In this work, we transform the conservation equations in cylindrical coordinates to spiral coordinates. The spiral coordinates are defined as $\psi = \phi + \beta(r)$. Here $\beta$ connects to the radial distances and spirality of the disk. Now, we consider the self-similarity conditions in the spiral coordinate as \citep{Chakrabarti-90b, Aktar-etal21}
\begin{align*} 
v_r &= r^{-1/2} q_{1}(\psi), \tag{7a}  \label{self_similar_a}\\
v_\phi &= r^{-1/2} q_{2}(\psi), \tag{7b} \label{self-similar_b}\\
a &= r^{-1/2} q_{3}^{1/2}(\psi), \tag{7c}\\
\rho &= r^{-3/2} q_{\rho}(\psi),  \tag{7d}\\ 
P &= r^{-5/2} q_{P}(\psi)  \tag{7e},
\end{align*}
and
\begin{align*}
\frac{\partial \beta}{\partial r} = r^{-1} B \tag{7f} \label{self_similar_f},
\end{align*}
where, `spirality', $B = \tan \theta$, and $\theta$ is the pitch angle (PA). 
The measure of entropy $K$ remains constant along the flow between two consecutive shocks; however, it changes at the shock. Here, $a$ represents the sound speed of the flow. Using the definition of sound speed, we calculate the variation of $K$ as
\begin{equation*}
K = r^{3\gamma/2-5/2}K_0 \tag{8}
\end{equation*}
where, $K_0 = \frac{q_P}{q_\rho}$ \citep{Chakrabarti-90a}. The entropy should generally increase inward for accretion and outward for wind. In this paper, we are interested only on the accretion solution. Therefore, we always choose $\gamma < 5/3$ to analyze accretion flow.

Now we obtain the disks height $(h)$ from equation (\ref{vertical_eq}) as
\begin{equation*}
h  = \frac{r^{-1/2} q_3^{1/2}}{\mathcal{G}}
\tag{9}
\label{height_eq}
\end{equation*}
where, $P = \rho a^2$ and $\mathcal{G} = \left(\frac{1}{r^3}  + \sigma \alpha r^{-3/2} \right)^{1/2}$. Here, $\alpha = 4 \pi q_\rho$. It is evident that the disk height is dependent on the density of the material present in the disk for the self-gravitating disk. Here, the surface density of the disk can be obtained as $\Sigma = 2 \rho h$. In general, if the disk is predominately dominated by disk gravity with negligible central object mass, the surface density follows $\Sigma \sim 1/r$ relation \citep{Bertin-Lodato99, Bertin-Lodato-01}. On the other hand, in a real spiral galactic disk, the surface density profile may be completely different, as depicted in equation \ref{height_eq}. In our present model, we consider a point-wise self-similar approach to incorporate spiral coordinate \citep{Aktar-etal21}. Our self-similarity model is valid point-wise, i.e., within a fixed radial distance ($r$). Therefore, it is difficult to infer the radial dependence of flow variables in the present formalism.

Therefore, we obtain the dimensionless differential equations of $q_1$, $q_2$ and $q_3$ from equations (\ref{radial_eq} - \ref{vertical_eq}) using equations (\ref{self_similar_a} - \ref{self_similar_f}) and (\ref{height_eq}), and are given by
\begin{equation*}\label{dimless_radial_eq}
 q_w \frac{d q_1}{d \psi} - \frac{n_\rho + 1}{\gamma} q_3 + \frac{ B}{(\gamma -1)} \frac{dq_3}{d\psi}  - \frac{q_1^2}{2} - q_2^2  + 1 
+   \frac{\sigma \alpha q_3^{1/2}}{\mathcal{G}} =0  ~~~~~~\tag{10}
\end{equation*}
\begin{equation*}\label{dimless_azimuthal_eq}
q_w \frac{d q_2}{d \psi} + \frac{ q_1 q_2}{2} + \frac{1}{(\gamma -1) }  \frac{d q_3}{d \psi}=0~~~~~~~~~~~~~~\tag{11}.
\end{equation*}
, and
\begin{align*}\label{dimless_vertical_eq}
& B \frac{d q_1}{d \psi} + \frac{d q_2}{d \psi} +  \frac{(\gamma +1)q_w}{2(\gamma -1) q_3} \frac{d q_3}{d \psi} - \frac{\sigma \alpha r^{-3/2}}{2G^2}  \frac{q_w}{(\gamma -1) q_3} \frac{d q_3}{d \psi}
\\ &  - \frac{3}{2} q_1 + \frac{3}{2} \frac{q_1 r^{-3}}{\mathcal{G}^2} + \frac{3}{4} \frac{q_1 \sigma \alpha r^{-3/2}}{\mathcal{G}^2}  = 0   \tag{12}
\end{align*}
where, $q_w = q_2 + Bq_1.$

\subsection{Sonic point analysis}

Here, we obtain the sonic point conditions by eliminating $\frac{dq_1}{d \psi}$ and $\frac{dq_2}{d \psi}$ from equation (\ref{dimless_vertical_eq}) using equation (\ref{dimless_radial_eq}) and (\ref{dimless_azimuthal_eq}), and is given by
\begin{align*} \label{grad_dq3_eqn}
\frac{d q_3}{d \psi} = \frac{N}{D}. \tag{13}
\end{align*}
where,
\begin{align*}
N & =     -\frac{(n_\rho +1) B q_3}{\gamma} - \frac{B q_1^2}{2} - Bq_2^2 + B + \frac{B \sigma \alpha q_3^{1/2}}{\mathcal{G}}   
\\ & + \frac{q_1 q_2}{2}  + \frac{3}{2} q_w q_1 - \frac{3}{2} \frac{r^{-3} q_1 q_w }{\mathcal{G}^2} - \frac{3}{4} \frac{\sigma \alpha r^{-3/2} q_1 q_w }{\mathcal{G}^2}   \tag{14}
\end{align*}
, and
$$
D = -\frac{B^2}{(\gamma -1)} - \frac{1}{(\gamma -1)} + \frac{q_w^2 (\gamma + 1)}{2(\gamma -1)q_3} - \frac{\sigma \alpha r^{-3/2}}{2\mathcal{G}^2}  \frac{q_w^2}{(\gamma -1) q_3}.
\eqno(15)
$$

During the accretion process into the compact object, the denominator $(D)$ at equation (\ref{grad_dq3_eqn}) becomes zero at some surfaces, known as sonic surface $\psi=\psi_c$. Simultaneously, the numerator $(N)$ also has to be zero at sonic surfaces to maintain the smooth solution \citep{Chakrabarti89}. The vanishing condition of denominator $D=0$ provides the sound speed at the sonic surface as
\begin{equation*}
q_{3c}  = \frac{q_w^2}{(B^2+ 1)} \frac{\Lambda}{2}
\tag{16}
\label{sound_sonic_eq}
\end{equation*}
where, $\Lambda = \left[(\gamma +1)- \frac{\sigma \alpha r^{-3/2}}{\mathcal{G}^2} \right]$.

In the presence of shock, the velocity component perpendicular to the shock is
\begin{equation*}
q_{\bot} = \frac{q_2 + B q_1}{ (B^2 + 1)^{1/2}},
\tag{17}
\label{vel_perpendicular_eq}
\end{equation*}
and velocity component parallel to the shock is given by
$$
q_{\parallel} = \frac{q_1 - B q_2}{ (B^2 + 1)^{1/2}}.
\eqno(18)
$$
The value of Mach number at the sonic surface is obtained as $ M_c = \left(\frac{q_{\bot}}{a}\right)_c = \sqrt{\frac{2}{\Lambda}}$ using equation (\ref{sound_sonic_eq}) and (\ref{vel_perpendicular_eq}). It is to be noted that the mach number at the sonic point deviates from the axisymmetric vertical equilibrium model for the self-gravitating disk.

On the other hand, the vanishing condition of numerator $N=0$ gives rise to the radial velocity $(q_{1c})$ at the sonic surface and is given by
$$
q_{1c} = \frac{- \mathcal{B} \pm \sqrt{\mathcal{B}^2 - 4 \mathcal{A} \mathcal{C}}}{2 \mathcal{A}}
\eqno(19)
$$
where,
\begin{align*}
\mathcal{A} & =  - \frac{B^3 (n_\rho + 1)}{(B^2 + 1)}  \frac{\Lambda}{2 \gamma} + B - \frac{3}{2}\frac{B r^{-3}}{\mathcal{G}^2} - \frac{3}{4} \frac{B \sigma \alpha r^{-3/2}}{\mathcal{G}^2}
\end{align*}
\begin{align*}
\mathcal{B} & = - \frac{ B^2 q_2 (n_\rho +1) }{(B^2 + 1)}  \frac{\Lambda}{\gamma} + \frac{B^2 \sigma \alpha}{\mathcal{G}}  \left[\frac{\Lambda}{2 (B^2 +1)} \right]^{1/2} + 2 q_2 
\\ & - \frac{3}{2}\frac{r^{-3}q_2}{\mathcal{G}^2} - \frac{3}{4} \frac{\sigma \alpha r^{-3/2}q_2}{\mathcal{G}^2}
\end{align*}
\begin{align*}
\mathcal{C} & = - \frac{B q_2^2  (n_\rho +1)}{(B^2 + 1)} \frac{\Lambda}{2 \gamma} - B q_2^2 + B + \frac{B \sigma \alpha q_2}{\mathcal{G}}  \left[\frac{\Lambda}{2 (B^2 +1)} \right]^{1/2}    \tag{20}   
\end{align*}
where the subscript, ``c'', represents the quantities evaluated at the sonic surface. To obtain the derivative $\frac{d q_3}{d \psi}|_c$ at the sonic surfaces, we apply `l'Hospital rule in equation (\ref{grad_dq3_eqn}) similar to \citet{Aktar-etal21}.

\begin{figure*}
	\begin{center}
		\includegraphics[width=0.90\textwidth]{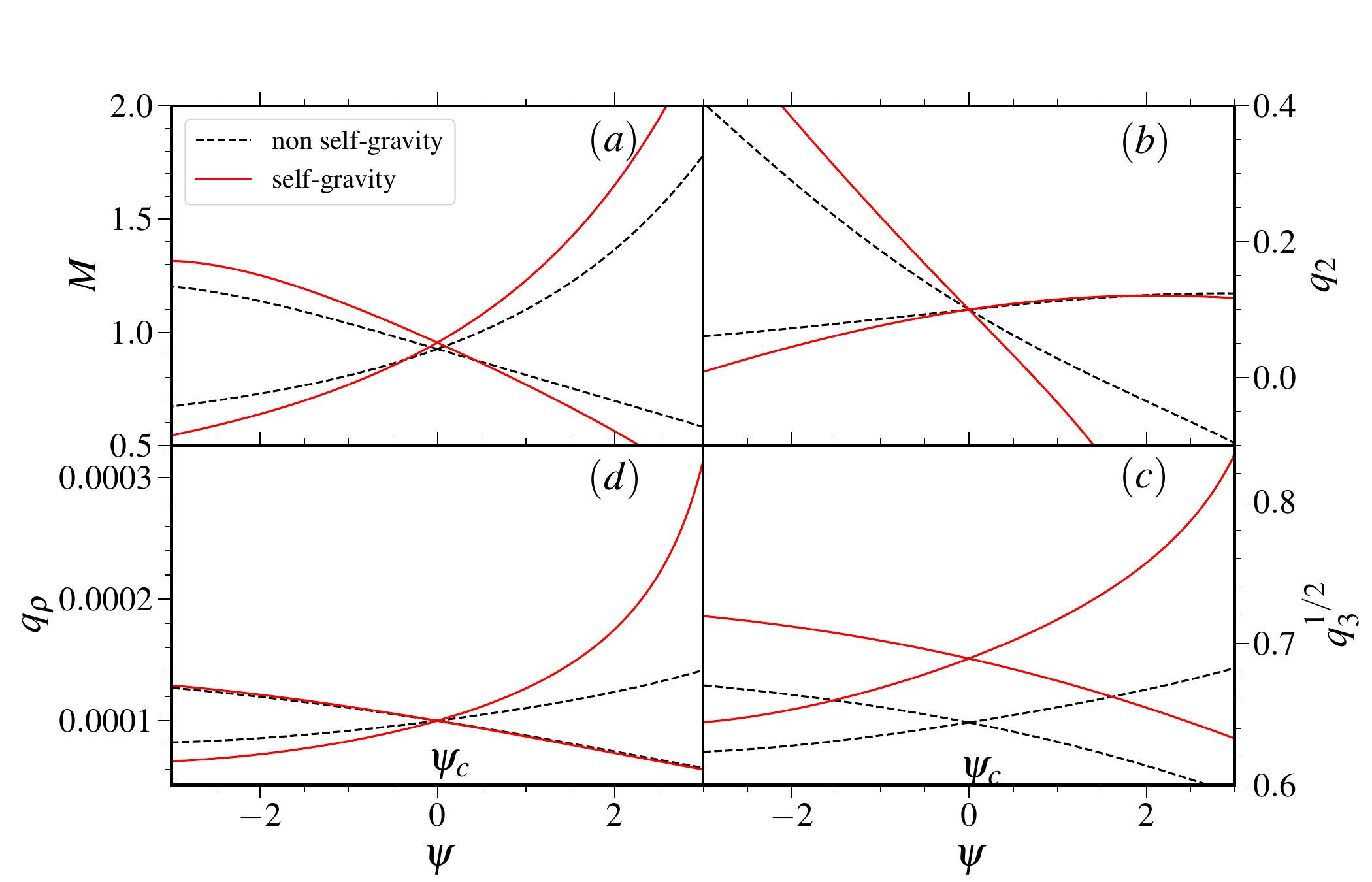} 
	\end{center}
	\caption{Comparison of flow variables for non self-gravitating and self-gravitating disk with 
 the spiral coordinates. The panel (a), (b), (c) and (d) represent mach number $(M)$, rotational 
 velocity $(q_2)$, sound speed or equivalently disk height $(q_3^{1/2})$, and density of flow 
 $(q_\rho)$ respectively. Here, we fix the flow parameters $(\theta, q_{2c}, q_{\rho c})$= 
 $(50^{\circ}, 0.10,10^{-4})$. For the calculation of self-gravitating disk, we fix the radial 
 distance at $r=25$. See the text for details. }
	\label{Figure_1}
\end{figure*}

\begin{figure}
	\begin{center}
		\includegraphics[width=0.50\textwidth]{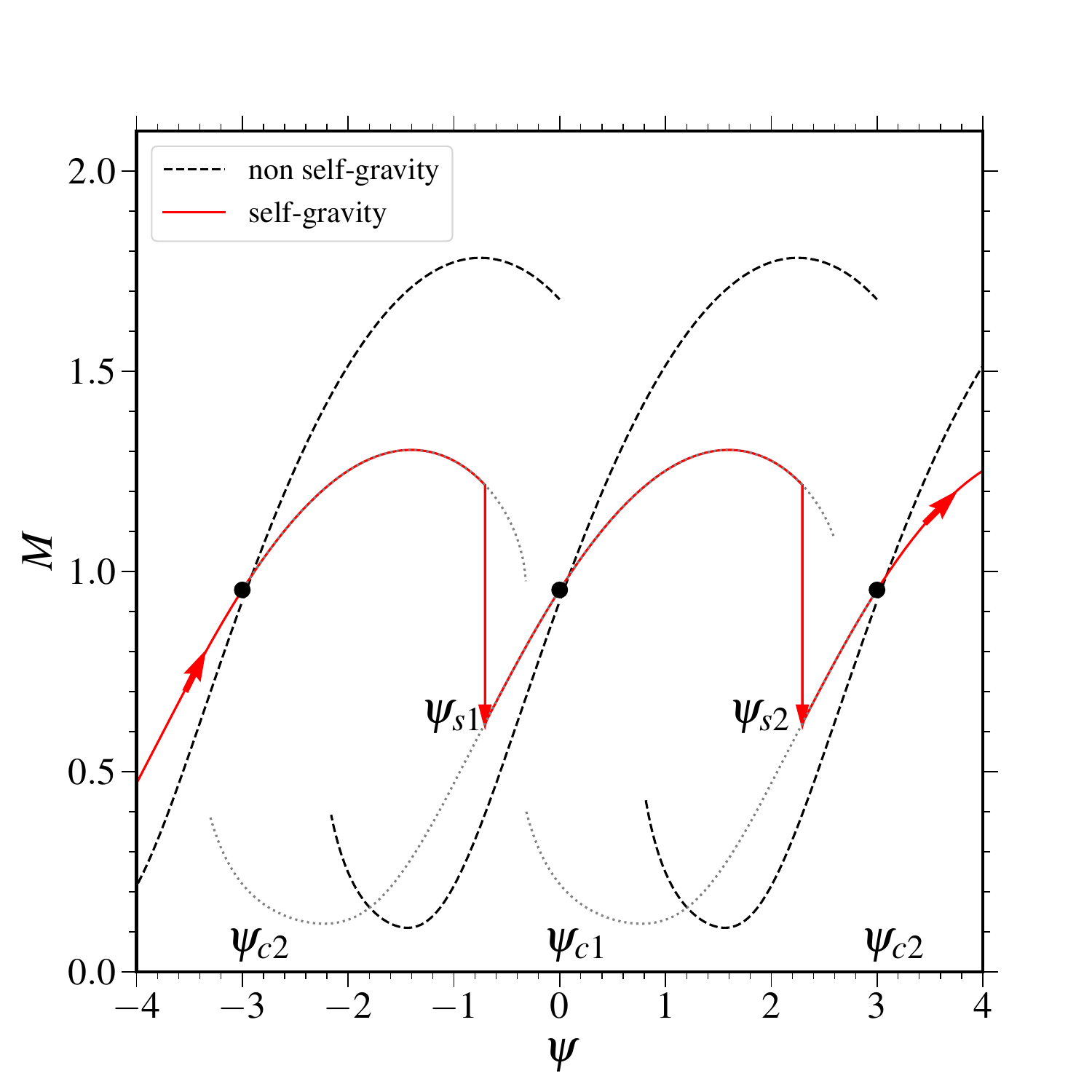} 
	\end{center}
	\caption{Comparison of solution topology for non self-gravitating and self-gravitating disk. The flow variables are $(\theta, q_{2c}, q_{\rho c})$= $(60^{\circ}, 0.86,10^{-4})$. We also fix $r = 25$. See the text for details. }
	\label{Figure_2}
\end{figure}

\begin{figure}
	\begin{center}
		\includegraphics[width=0.50\textwidth]{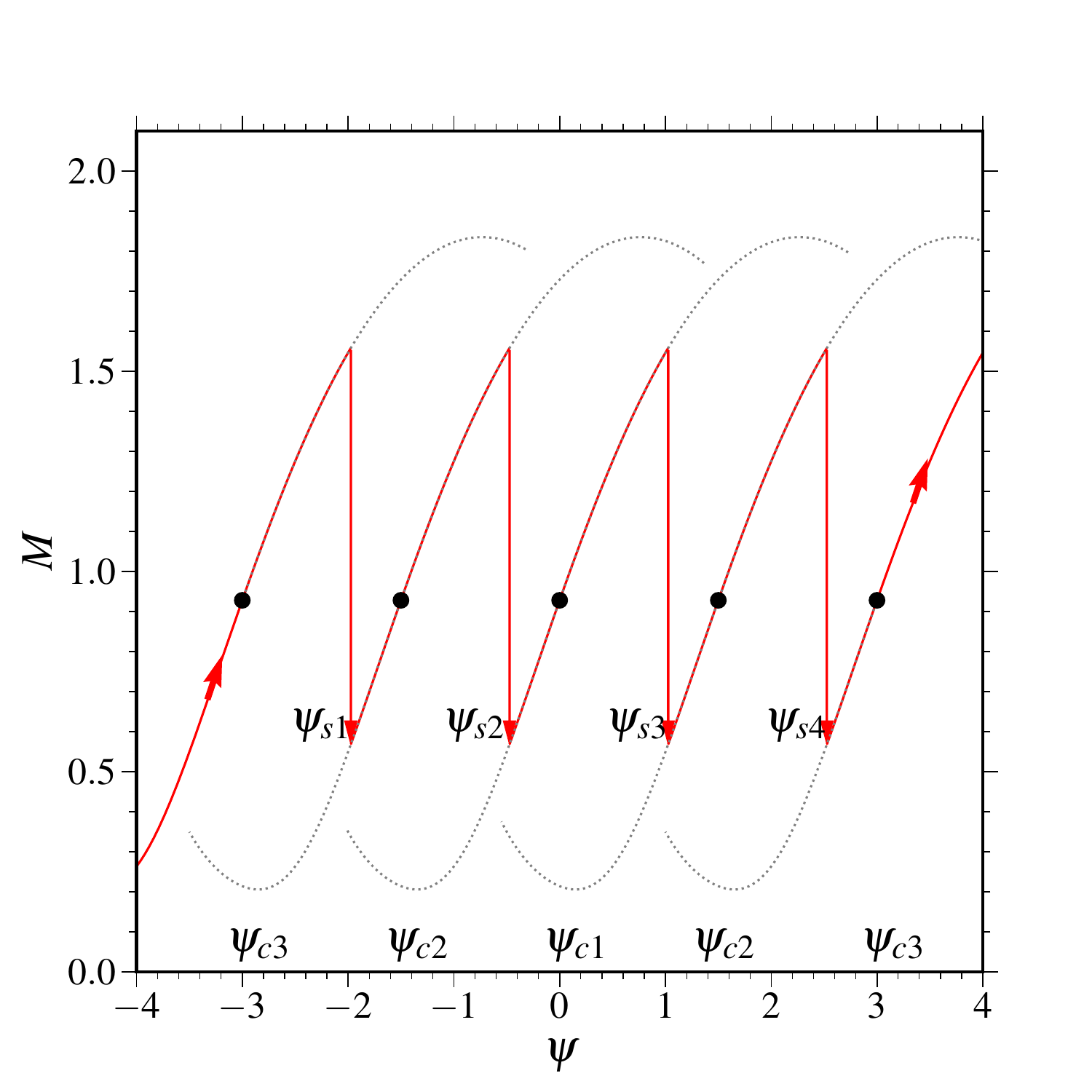} 
	\end{center}
	\caption{Representation of spiral shocks transitions for the number of shocks $n_s =4$ in presence of self-gravitating disk. The vertical arrows represent spiral shock transitions in the flow. Here, the  flow parameters are $(\theta, q_{2c}, q_{\rho c}, r)$= $(45^{\circ}, 0.75,10^{-5}, 20)$. See the text for details.}
	\label{Figure_3}
\end{figure}

\begin{figure}
	\begin{center}
		\includegraphics[width=0.48\textwidth]{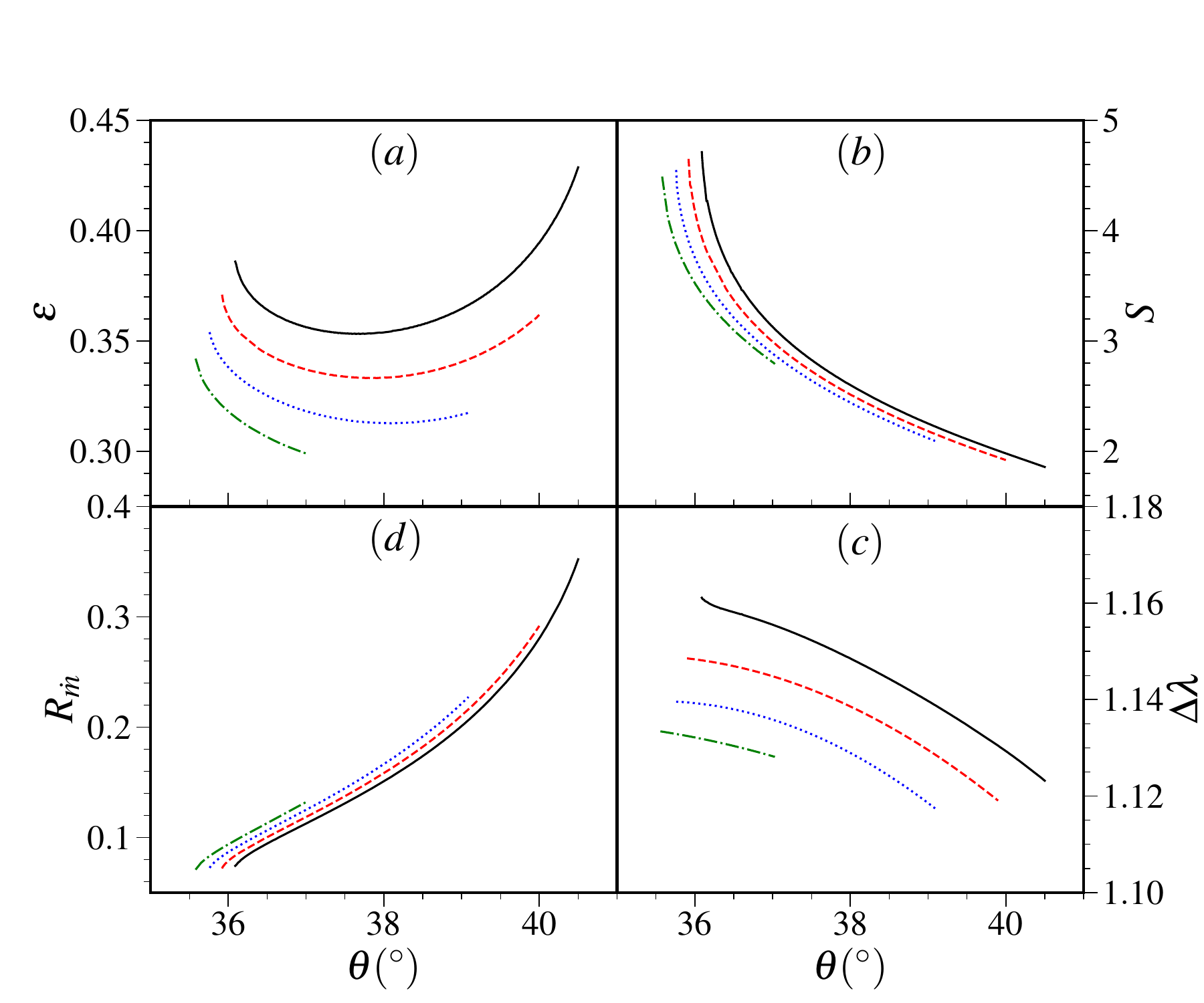} 
	\end{center}
	\caption{Variation of (a): shock locations ($\epsilon$), (b): shock strength ($\mathcal{S}$), (c): amount of angular momentum dissipation ($\Delta \lambda$) across shock, and (d): mass outflow rate ($R_{\dot{m}}$) in terms of pitch angle for various flow density at sonic surface ($q_{\rho c}$). The solid (black), dashed (red), dotted (blue), and dashed-dotted (green) curves are for $q_\rho = 0.0001, 0.1, 0.2$ and 0.3 respectively. Here, we fix $q_{2c} = 0.60$ at the radial distance $r = 0.01$. See the text for details.}
	\label{Figure_4}
\end{figure}

\begin{figure}
	\begin{center}
		\includegraphics[width=0.48\textwidth]{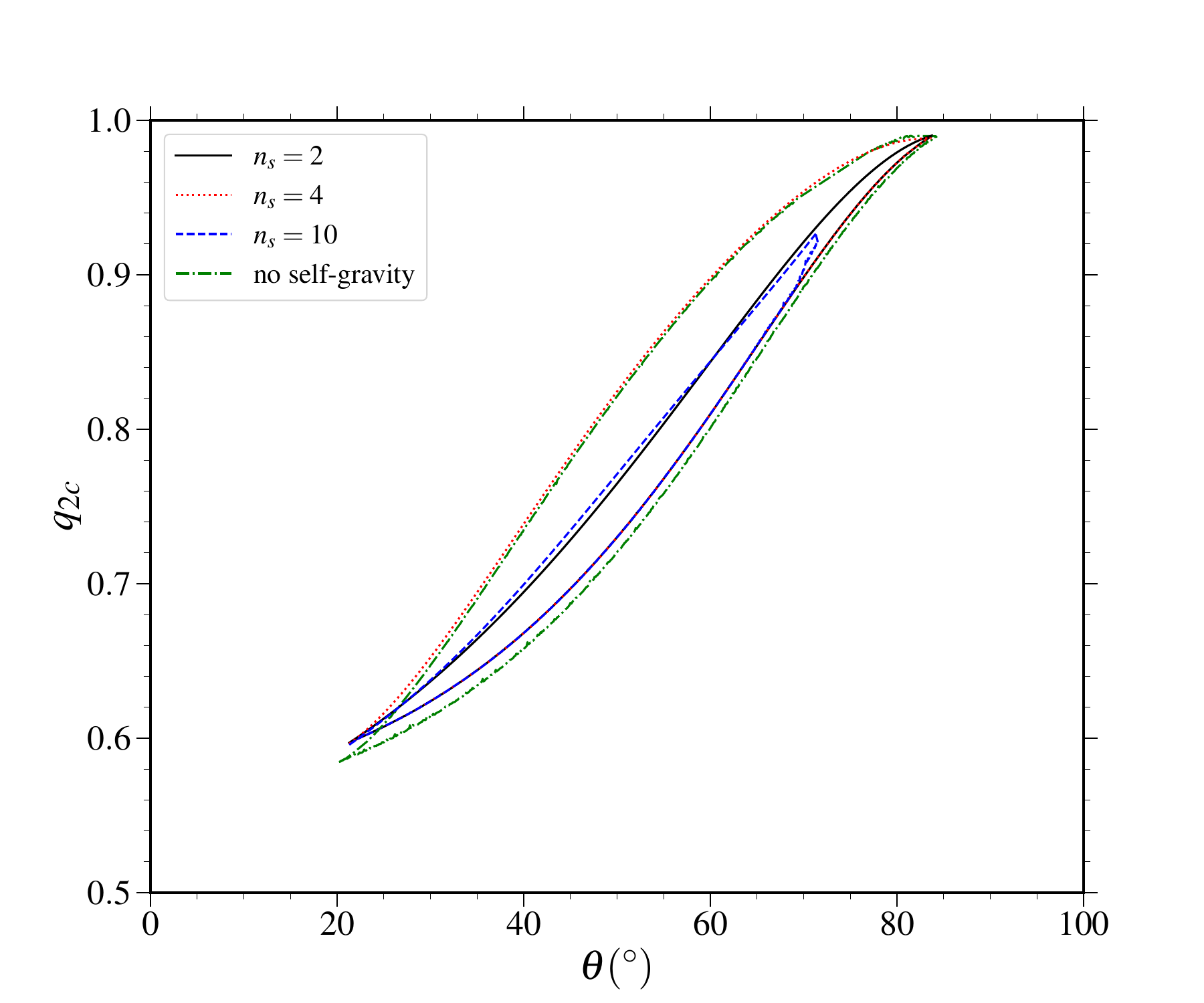} 
	\end{center}
	\caption{Parameter space for different number of spiral shocks $(n_s)$. Here, we fix the radial distance at $r=10$ and the flow density as $q_{\rho c} = 10^{-4}$. See the text for details.}
	\label{Figure_5}
\end{figure}

\subsection{Computation of mass outflow rate from the disk}
\label{outflow_rates}

The net mass flux can be obtained from equation (\ref{continuity_eq}) using self-similar conditions (equations \ref{self_similar_a}-\ref{self_similar_f}). One part of the mass flux is contributed to radial inflow mass flux ($\dot{M}_{\rm in}$) (i.e., accretion rate), and another part is contributed to the wind flux in the azimuthal direction \citep{Chakrabarti-90b}. It is notable to mention that \citet{Chakrabarti-90b} did not consider the situation of mass outflow from the disk. In general, if there is no spiral shock in the flow, the wind flux is zero \citep{Chakrabarti-90b}. However, in the presence of spiral shocks, the post-shock matter is very hot and dense. Due to the excess thermal gradient force across shocks may drive the matter as mass outflow in the spiral arm from the disk similar to the axisymmetric accretion disk model \citep{Aktar-etal15, Aktar-etal17}. It is to be mentioned that here we assume a two-dimensional vertical equilibrium model (i.e., 2.5D). Therefore, estimating the mass flux component in the vertical direction is impossible in the present model. However, we argue that the mass flux in the azimuthal direction accumulates in the spiral arm and is ejected away as mass outflow from the disk due to the thermal gradient force across spiral shock waves. Now, in our model, if we consider mass outflow, we need to balance non-zero mass flux in azimuthal direction with mass outflow rates $\dot{M}_{\rm out}$ to maintain mass conservation equation (\ref{continuity_eq}).

Therefore, the mass accretion rate in the radial direction can be obtained from equation (\ref{continuity_eq}) as 
\begin{equation*}
\dot{M}_{\rm in} = \int_0^{2 \pi}{q_1 q_{\rho}q_3^{1/2}} d\psi.
\tag{21}
\label{mass_acc_eqn}
\end{equation*}
On the hand, the mass outflow rates from the disk are obtained by equating the wind flux normal to the spiral shock to preserve the total mass flux in the flow, i.e.,
\begin{equation*}
\dot{M}_{\rm out} \equiv \int_0^{2 \pi}{h q_{\rho} q_{w }} d\psi.
\tag{22}
\label{mass_out_eqn}
\end{equation*}
Here, the ratio of mass outflow to inflow rates can be calculated as $R_{\dot{m}}= \frac{\dot{M}_{\rm out} }{\dot{M}_{\rm in} }$ \citep{Aktar-etal15, Aktar-etal17, Aktar-etal21}.


\subsection{Spiral shock conditions and solution methodology}

The spiral shock conditions are given by \citep{Chakrabarti-90b, Aktar-etal21}

(1) The energy conservation:
\begin{align*} \label{shock_condition1}
\frac{q_{3+}}{\gamma -1} + \frac{q_{\bot +}^2}{2} = \frac{q_{3-}}{\gamma -1} + \frac{q_{\bot -}^2}{2}
\tag{23a}
\end{align*}

(2) The momentum conservation:
\begin{align*}\label{shock_condition2}
W_+ + \Sigma_{+} q_{\bot +}^2 = W_- + \Sigma_{-} q_{\bot -}^2
\tag{23b}
\end{align*}

(3) The conservation of mass flux normal to the shock:
\begin{align*}\label{shock_condition3}
h_+ ~q_{\rho +}~ q_{w +} = h_-~ q_{\rho -}~ q_{w -}
\tag{23c}
\end{align*}

(4) The conservation of velocity component parallel to the flow:
\begin{align*} \label{shock_condition4}
q_{1+} - Bq_{2+} = q_{1-} - Bq_{2-}
\tag{23d}
\end{align*}
where ``$\pm$'' implies post-shock and pre-shock quantities, respectively. Here, $W$ represents the vertically integrated gas pressure of the flow \citep{Matsumoto-etal84, Chakrabarti89}. The shock invariant quantity $(C_s)$ is obtained using equations (\ref{shock_condition1}-\ref{shock_condition3}) as
\begin{align*} \label{shock_invariant}
C_s = \frac{\left[M_{+} (3\gamma -1) + \frac{2}{M_+}\right]}{\left[2 + (\gamma -1)M_{+}^2\right]} = \frac{\left[M_{-} (3\gamma -1) + \frac{2}{M_-}\right]}{\left[2 + (\gamma -1)M_{-}^2\right]}.
\tag{24}
\end{align*}
We define the shock strength as $\mathcal{S} =\frac{M_-}{M_+}$.
 The analytical expression of shock location $\epsilon$ can be obtained as \citep{Chakrabarti-90b, Aktar-etal21}, and is given by
\begin{align*}  \label{shock_location}
\epsilon = \frac{1}{2} + \frac{1}{\delta \psi} \frac{\left(\frac{d q_{\parallel}}{d \psi}\right)_c}{\left(\frac{d^2 q_{\parallel}}{d \psi^2}\right)_c},
\tag{25}
\end{align*}
where, $\delta \psi = 2\pi/n_s$. $n_s$ is the number of shocks in the flow. We obtain the calculation of the second-order derivatives at the sonic surfaces in a similar way of \citet{Aktar-etal21}. We ignore to represent the long expression here to avoid repetition. Further, we quantify the amount of specific angular momentum $(\lambda)$ dissipated in the presence of the spiral shocks as
\begin{align*} \label{ang_diss}
\Delta \lambda = \frac{\lambda_+}{\lambda_-} = \frac{q_{2+}}{q_{2-}}. \tag{26}
\end{align*}

The classical self-similar solution is a common feature for the Newtonian gravitational potential, and it has been widely investigated in literature starting with some pioneering works \citep{Spruit-87, Chakrabarti-90b, Narayan-Yi94}. The self-similar solutions make flow equations dimensionless and independent of position. This approach can be widely applied in various physical situations in accretion physics. However, the classical self-similar approach unable to incorporate various interesting physical scenarios, such as the non-inertial effects from the co-rotating frame of the binary, self-gravitating disk, etc. Moreover, the numerical simulations also indicate that the self-similar solution is only valid in the middle radial region of the accretion disk in which there is less effect from
the inner and outer boundaries \citep{Narayan-Yi94}. In general, it is pointed out that the self-similar solution is only a local solution under local simplification but
not a global solution. Recently, \citet{Aktar-etal21} considered the self-similar condition to simplify the calculation and obtain the point-wise valid solution to incorporate the physical effects from the companion gravity, centrifugal force, and Coriolis’ force. Motivating by this, we also adopt point-wise self-similar solutions to investigate spiral shocks in a self-gravitating disk by incorporating self-gravitating potential in our model.

Here, we adopt the same solution methodology, i.e., the point-wise self-similar approach proposed by \citet{Aktar-etal21}. We first fix the radial distance $(r)$ of the flow. Then, to obtain the solution, we apply the same input parameters mentioned by \citet{Chakrabarti-90b}. Therefore, we supply the number of shocks $(n_s)$, pitch angle $(\theta)$, rotational velocity at the sonic surface $(q_{2c})$, and adiabatic index ($\gamma$) of the flow. Additionally, we need to supply the density $(q_{\rho c})$ at the sonic surface due to the consideration of a self-gravitating disk. We self-consistently determine the shock location $(\epsilon)$ using equations (\ref{shock_invariant}) and (\ref{shock_location}). We also fix the adiabatic index $\gamma = 4/3$ throughout the paper; otherwise, it is stated.

\section{Results}
\label{results}

In a non-axisymmetric accretion flow, the inflowing matter spirals around the compact object. During accretion, the flow might encounter several spiral shock transitions depending on the flow parameters \citep{Chakrabarti-90b, Aktar-etal21}. Due to the shock transition, the flow losses its angular momentum (see equation \ref{ang_diss}) and enters into the central compact object. If the gravitational field due to the matter present in the disk is significant enough, we need to incorporate the self-gravitating effect in governing equations. Keeping this in mind, we consider the self-gravitating effect in our present formalism. To obtain the solutions, we first need to examine the nature of the sonic surfaces, which can be determined following the quadratic expression of $\frac{d q_3}{d \psi}|_c$ at the sonic surfaces. The sonic surfaces can be broadly classified into two ways, such as physical (discriminant > 0) and unphysical (discriminant < 0) sonic surfaces. The physical sonic surfaces are also classified into `saddle type,' `straight line', and `nodal type' depending on various conditions (see \citet{Aktar-etal21, Chakrabarti-90a} for details). In this work, we first identify the saddle-type sonic surfaces by supplying the inflow parameters, namely pitch angle ($\theta$), the rotational velocity at the sonic surface ($q_{2c}$) and density of matter $(q_{\rho c})$ at the radial position $(r)$, respectively \citep{Chakrabarti89, Chakrabarti-90b}. In order to obtain solutions, we numerically integrate equation (\ref{dimless_radial_eq}-\ref{dimless_vertical_eq}) from saddle type sonic surfaces by supplying the flow variables $(\theta, q_{2c}, q_{\rho c}, \gamma)$ using both the slope of $\frac{d q_3}{d \psi}|_c$ at a particular radial distance $(r)$ (see \citet{Aktar-etal21}). During accretion, the flow passes through spiral shocks, and due to the shock compression, a part of the accreting matter emerges from the disk as mass outflows \citep{Aktar-etal15, Aktar-etal17, Aktar-etal21}. Here, the mass outflow rates are calculated using equations (\ref{mass_acc_eqn}-\ref{mass_out_eqn}). To begin with, we first investigate the comparison between non self-gravitating and self-gravitating disk accretion flow. For the purpose of comparison, we investigate mach number $(M)$, rotational velocity $(q_2)$, sound speed $(q_3^{1/2})$, and density of the matter $(q_\rho)$ with the spiral coordinates, respectively. In panel $(a)$ of Figure \ref{Figure_1}, we compare the mach number with spiral coordinates $(\psi)$. We observe that mach number variation is completely different for the self-gravitating disk compared to the non self-gravitating disk. Moreover, the mach number at the sonic point $(M_c)$ is different from the non self-gravitating disk as indicated in equations (\ref{sound_sonic_eq}-\ref{vel_perpendicular_eq}), depicted in Figure \ref{Figure_1}a. In a similar manner, the rotational velocity $(q_2)$ and the sound speed or equivalent disk height $(q_3^{1/2})$ varies significantly in the presence of self-gravitating disk, shown in Figure \ref{Figure_1}b and \ref{Figure_1}c, respectively. It is obvious that the disk height increases in a particular radial position with the spiral coordinates for self-gravitating disk compared to non self-gravitating disk, depicted in Figure \ref{Figure_1}c. On the other hand, we also observe that the density of matter $(q_\rho)$ also deviates from the non self-gravitating disk even starting from the same sonic values, as shown in Figures \ref{Figure_1}d. Here, we fix the flow parameters $(\theta, q_{2c}, q_{\rho c})$ as $(50^\circ, 0.10, 10^{-4})$ at the radial distance $r = 25$. The solid (red) and dashed (black) curves are for self-gravitating and non self-gravitating disks, respectively. The saddle-type sonic surfaces $(\psi_c)$ are indicated in the figure. It is to be noted that the particular solution mentioned in Figure \ref{Figure_1} does not exhibit spiral shocks.

Now we investigate the comparison of solution topology between non self-gravitating and self-gravitating disks in Figure \ref{Figure_2} for $n_
s =2$. During accretion, the inflowing matter passes through some sonic surfaces $(\psi_c)$ to become supersonic. If the spiral shock conditions (equations \ref{shock_condition1} - \ref{shock_condition4}, \ref{shock_invariant}, \ref{shock_location}) are satisfied, then the flow makes a discontinuous jump to the subsonic flow. Immediately, the flow picks up its velocity and again passes through another sonic surface. Again the shock transition happens, and the flow loses its angular momentum. Finally, the matter enters into the compact object. The vertical arrows indicate the spiral shock transitions in the flow and solid (black) circles represent sonic surfaces. Here, the solid (red) and dashed (black) curves are represented for self-gravitating and non self-gravitating disks, respectively. Interestingly, we find two spiral shocks $(n_s =2)$ for self-gravitating disk, albeit there are no spiral shocks present in non-self-gravitating flow for the same inflow parameters. The corresponding sonic surfaces ($\psi_{1c}, \psi_{2c}, \psi_{3c}$) and shock locations ($\psi_{s1}, \psi_{s2}$) are also indicated in the figure. Here, the shock parameters are $(\epsilon, \mathcal{S})$ = ($0.2700, 1.9526$). In a similar way, we also present solution topology in the presence of a self-gravitating disk when the number of spiral shocks is $n_s = 4$, depicted in Figure \ref{Figure_3}. The corresponding four sonic surfaces $(\psi_{1c}, \psi_{2c}, \psi_{3c})$, and four shock locations ($\psi_{s1}, \psi_{s2}, \psi_{s3}, \psi_{s4}$) are shown in the figure. The corresponding shock parameters are $(\epsilon, \mathcal{S})$ = ($0.3469, 2.7238$). We fix the flow parameters as $(\theta, q_{2c}, q_{\rho c}, r)$ as $(60^\circ, 0.86, 10^{-4}, 25)$ and $(45^\circ, 0.75, 10^{-5}, 20)$ for Figure \ref{Figure_2} and Figure \ref{Figure_3}, respectively.

Further, we examine the overall behavior of shock properties in terms of pitch angle by fixing all other flow parameters. In Figure \ref{Figure_4}a, we represent shock location $(\epsilon)$ with the variation of pitch angle $(\theta)$. Here, solid (black), dashed (red), dotted (blue), and dashed-dotted (green) are for different flow densities at sonic surface $q_{\rho c} = 0.0001, 0.1, 0.2$, and 0.3, respectively. The corresponding shock strength $(\mathcal{S})$ is plotted in Figure \ref{Figure_4}b. We observe that shock strength decreases with the increase of pitch angle. This implies that a tighter spiral arm exhibits stronger spiral shocks in the flow. Also, shock strength decreases with the increase of density of the flow for a particular pitch angle. Similar trends have been observed for dissipation of angular momentum $(\Delta \lambda)$ as shock strength, depicted in Figure \ref{Figure_4}c. On the other hand, the mass outflow rates $(R_{\dot{m}})$ is plotted in Figure \ref{Figure_4}d. It is found that mass outflow rates increase with the pitch angle. It clearly indicates that gaseous matter can escape more easily from the disk due to spiral shocks for a weakly wound spiral arm compared to the strong one. Also, mass outflow rates are higher for a denser flow than a less dense flow for a particular pitch angle due to the availability of more matter in the disk surface. Here we fix rotational velocity $q_{2c} = 0.6$ at the radial distance $r = 0.01$.   

So far, we have compared the solution topology. Now, we investigate the overall parameter space containing spiral shocks. The shock parameter space is separated by pitch angle $(\theta)$ and rotational velocity $(q_{2c})$ at sonic surfaces, shown in Figure \ref{Figure_5}. Theoretically, the number of spiral shocks lies within the range $1\geq n_s\rightarrow \infty$ \citep{Spruit-87}. Therefore, we compare the parameter space for a self-gravitating disk for the number of shocks $n_s =2, 4$, and 10. Here, we fix the radial distance at $r = 10$ and density at sonic surface $q_{\rho c} = 10^{-4}$. We observe that shock parameter space increases significantly from $n_s =2$ to $n_s =4$, decreasing again for $n_s =10$. There is a clear indication that the parameter space shrinks for the higher number of shocks $n_s$. It also indicates that the shock solutions are less probable for the higher number of shock solutions. Along with that, we also plot the parameter space for non self-gravitating disk, which is independent of radial position \citep{Chakrabarti-90b}. Here, we choose the number of shocks $n_s =4$. In Figure \ref{Figure_5}, solid (black), dotted (red), and dashed (blue) curves are for self-gravitating disks for the number of shocks $n_s =2, 4$, and 10, respectively. The corresponding dashed-dotted (green) curve is for the non self-gravitating disk. This parameter space of non self-gravitating disk is the same as Figure 6 of \citet{Chakrabarti-90b} for accretion solution (i.e., $\sigma =0)$.

\section{Application to spiral galactic gaseous disks}
\label{application}

In this section, we apply our model to spiral galactic gaseous disks. The shock compression, due to spiral shocks, drives some of the shocked gaseous matter forming star, while some directly outflow the galactic disk. However, the detailed physical processes of star formation will depend on the specific galactic environment. In our present work, inferring the galactic environment and various other physical processes is impossible. Two parameters, namely PA and shear rate (SR), both play pivotal roles in spiral galaxy properties. We assume that the spiral shock wave and the spiral arm have the same PA since they are always associative; however, they are actually a little different. Therefore, here we consider the PA estimated by the galaxy image analysis as the PA of the spiral shock wave. For the SR ($\Gamma$), its definition can easily be found in previous works (e.g., \citet{Seigar-etal05, Seigar-etal06, Yu-Ho19}) and is given as
\begin{align*} \label{theta_vs_Gamma-1}
\Gamma=1-\frac{r}{V_c}\frac{dV_c}{dr}, \tag{27}
\end{align*}
where $r$ and $V_c$ are the radial distance and the circular velocity around the galactic center, respectively. In our model, the azimuthal velocity $v_{\phi}$ is identical to the $V_c$ in equation (\ref{theta_vs_Gamma-1}). Therefore, replacing $V_c$ with the equation (\ref{self-similar_b}) and doing the mathematical simplification, we obtain from the equation (\ref{theta_vs_Gamma-1})
\begin{align*}\label{theta_vs_Gamma-2}
\Gamma=\frac{3}{2}-k\tan\theta, \tag{28}
\end{align*}
where $k\equiv d\ln q_2/d\psi$, which is one of the derivatives in our model (see equation \ref{dimless_azimuthal_eq}) and represents the changing rate of azimuthal velocity along the $\psi$-direction. It is an interesting fact that $\Gamma$ is equal to $3/2$ if the Keplerian velocity $V_c=\sqrt{GM/r}$ is substituted into equation (\ref{theta_vs_Gamma-1}), which implies the effect of the mass $M$ is only concentrated inside the radius $r$ (no mass outside $r$). Therefore, on the right side of the equation (\ref{theta_vs_Gamma-2}), $3/2$ is the Keplerian upper limit of SR, and the term of $k\tan\theta$ represents the effects from the disk mass and dark matter halo.  

The PA, $\theta$, can be measured from the galaxy images through the discrete Fourier transformation, and the SR, $\Gamma$, can be estimated by fitting the circular velocity curve (CVC) of the galaxy. \citet{Seigar-etal05, Seigar-etal06} supplied a set of measuring PAs and SRs for a total of 45 galaxies by the images in the near-infrared/optical band and the observational CVCs, respectively. Recently, \citet{Yu-Ho19} also provided a new data set including 79 galaxies, whose PAs were measured from their optical images of the Sloan Digital Sky Survey (SDSS) and SRs came from \citet{Kalinova-etal17} using the CVCs of the Calar Alto Legacy Integral Field Area (CALIFA). 

\begin{figure}
	\begin{center}
		\includegraphics[width=0.50\textwidth]{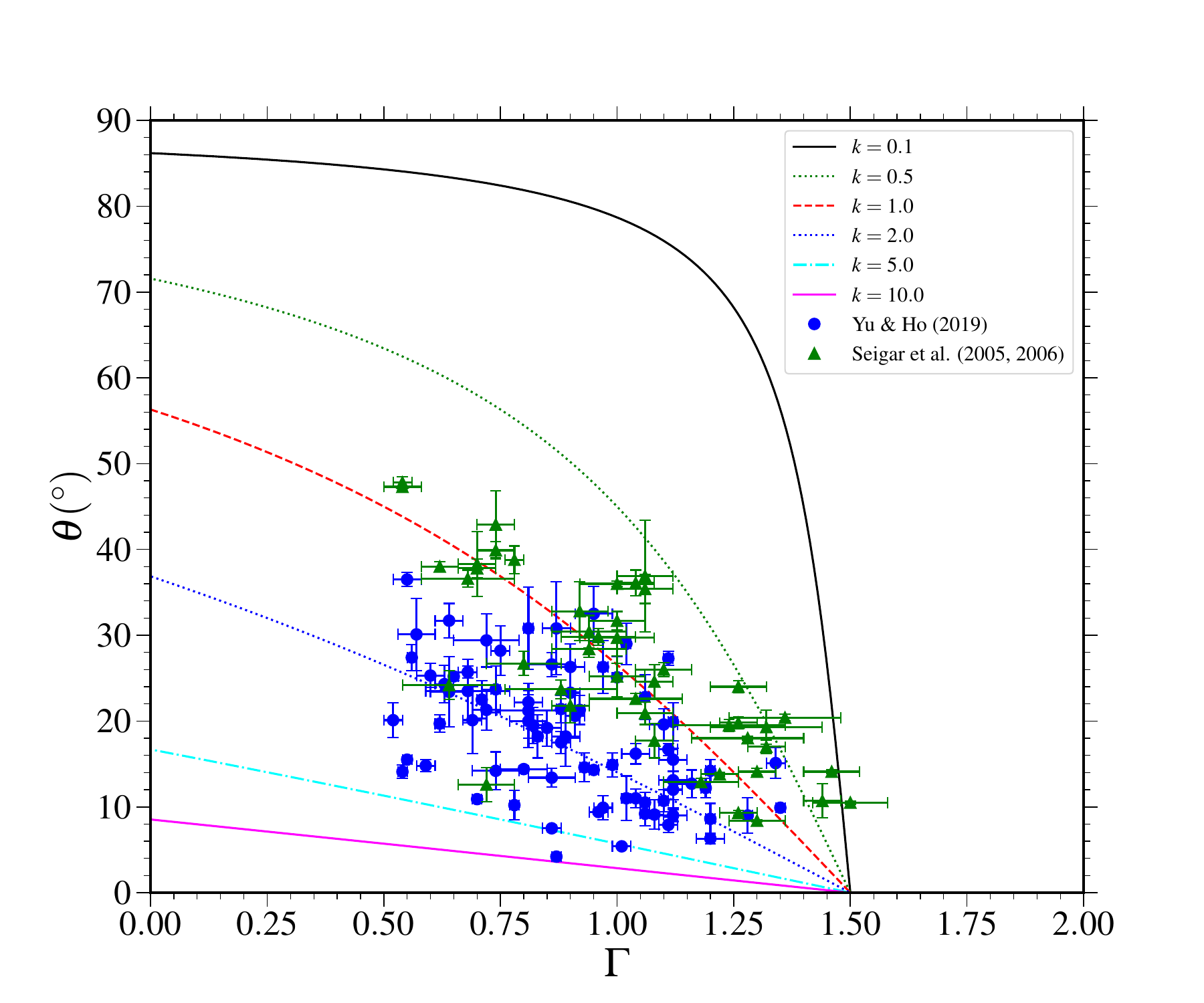} 
	\end{center}
	\caption{Theoretical value of $k$ in terms of pitch angle and shear rate. The blue circles and green triangles are for observational data of \citet{Yu-Ho19} and \citet{Seigar-etal05, Seigar-etal06}, respectively. See the text for details.}
	\label{Figure_6}
\end{figure}

In Figure \ref{Figure_6}, we represent a plot of $\theta$ vs. $\Gamma$, which contains the contours of $k$ defined in Equation (\ref{theta_vs_Gamma-2}) and the data points of 124 galaxies collected from \citet{Seigar-etal05, Seigar-etal06} (green triangles) and \citet{Yu-Ho19} (blue circles). It can be easily seen that most of the data points (green and blue) are distributed in the area between two curves of $k=0.5$ and $5.0$, and the dispersion of PA gradually contracted along the contour lines of $k$ when the SR increases to approach the limit of $3/2$. This interesting contractivity shows the intrinsic physical properties of this dispersion, which can be measured by $k$ with Equation (\ref{theta_vs_Gamma-2}). We can calculate $k$ from the observational PA and SR on a specific galaxy case and fix the derivate $d\ln q_2/d\psi$ in our model to constrain the property of spiral shocks coherent to spiral arms in this galaxy.  

Encouraging by this, we continue to analyze the correlations between the SFR and PA. In order to achieve this goal, we first need to calculate the physical quantities in a proper unit system. Hereafter, we use the unit system of $G=V_{\rm{K}}=r=1$ instead of $G=M=c=1$, which is used in the earlier part of this paper. Where the $V_{\rm{K}}\equiv\sqrt{GM/r}$ is the Keplerian velocity at location $r$ and the mass $M$ includes all of the visible and invisible mass, such as the gas, dust, star, and dark matter, etc., inside the radius $r$. In order to estimate $V_{\rm{K}}$ for a particular galaxy, we assume that the stars are approximately rotating with the local Keplerian velocity (It should be noted that the $v_{\phi}$ in our model is the circular velocity of the gas. Because our model is a hydrodynamic model, $v_{\phi}$ is not necessarily Keplerian and is scaled with the local Keplerian velocity). Then, we can obtain this local star rotating velocity at the location $r$ by interpolation from the CVC of this galaxy. Based on the principal component analysis (PCA) of \citet{Kalinova-etal17}, we can obtain this physical quantity accurately and conveniently by their PCA interpolating formulae. As mentioned earlier, our model solution is point-wise valid; our analysis is restrained locally. Therefore, the location $r$ refers to the radial distance from the galaxy's center to where these analyses and measurements occur. In this regard, we estimate $r$ as the radius of the middle point of the range used by Fourier analysis for the PA, i.e., $r=r_{\rm{mid}}$, which is the most probable measuring location of PA. \citet{Yu-Ho19} provided the radial ranges in unit arcsec for their 79 galaxies and we can obtain the distances from earth to these galaxies in unit Mpc from the data set of CALIFA, so we can finally calculate the $r_{\rm{mid}}$ in unit parsec, which is listed in Table \ref{Table-1} (column 6) for every galaxy from \citet{Yu-Ho19}. Moreover, we need to fix the number of spiral shocks $n_s$ for our model. Here, we assume the Fourier mode used to calculate pitch angle in \citet{Yu-Ho19} paper as the number of shocks (see column 4 of Table \ref{Table-1}). Unfortunately, we have not found any observational distance for the 45 galaxies analyzed by \citet{Seigar-etal05, Seigar-etal06}, so we cannot continue to use their sample for subsequent analysis in this paper.

\begin{figure}
	\begin{center}
		\includegraphics[width=0.50\textwidth]{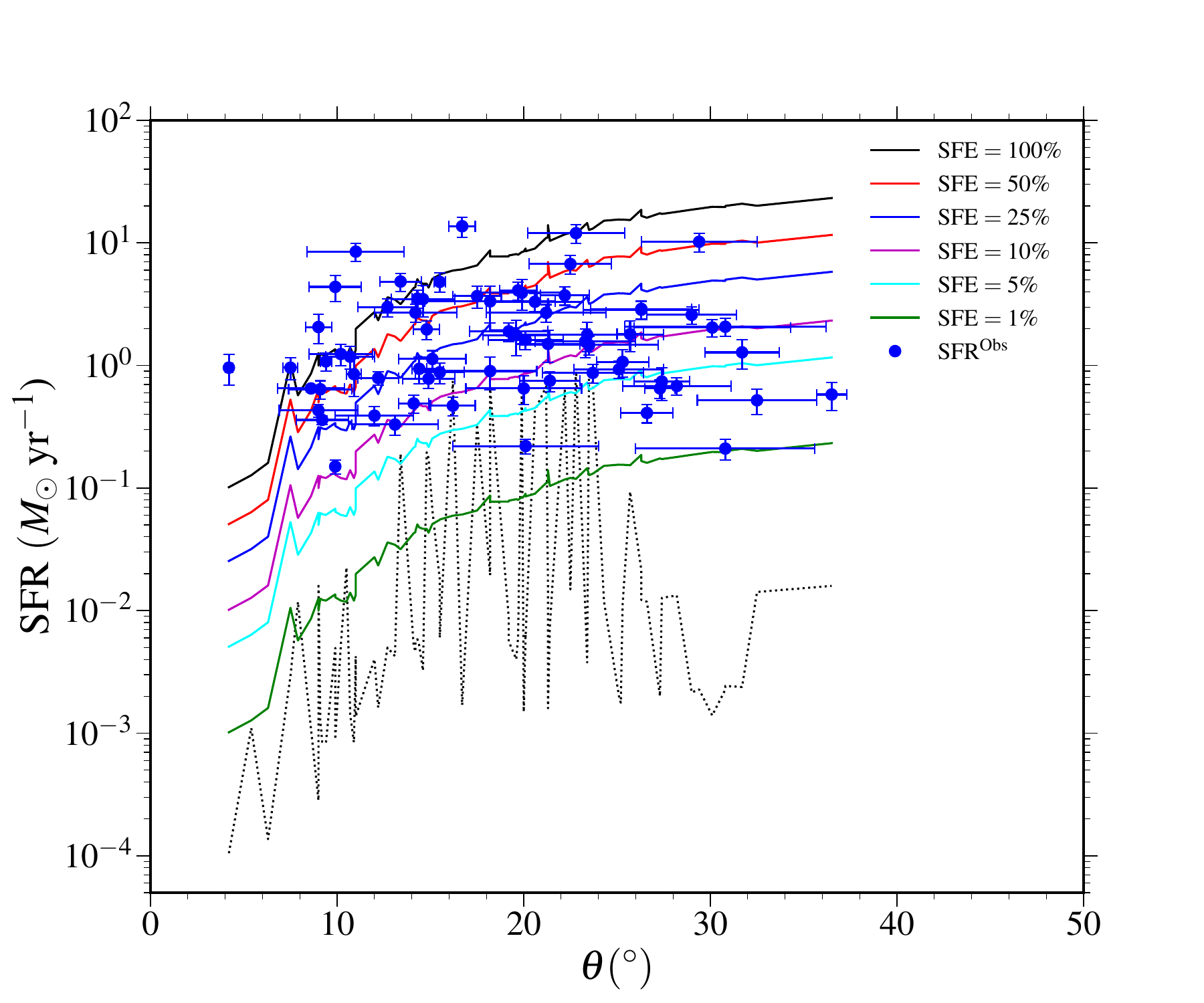} 
	\end{center}
	\caption{Model calculated maximum and minimum star formation rate (SFR) in terms of pitch angle $(\theta)$ for 79 spiral galaxies. Black solid and dotted curves are for the maximum and minimum of theoretically estimated SFR at SFE=100\%. The other coloured solid curves are for the maximums SFR at different SFEs.  The solid blue circles represent observed SFR (SFR$^{\rm Obs}$) \citep{Catalan-Torrecilla-15}. See the text for details.}
	\label{Figure_7}
\end{figure}

Next, we continue to estimate the SFR based on our model. As mentioned in the above Section \ref{outflow_rates}, we can calculate the outflow rate (equation \ref{mass_out_eqn}), which is regarded as the estimation of SFR induced by spiral shocks (arms) in our model. However, we are still unable to determine this SFR for each specific galaxy because there are two parameters ($q_{2c}$ and $q_{\rho c}$) undetermined. These two parameters represent the circular velocity and density of flow at the sonic surface, whose values might depend on the galactic ambient conditions at the shock front, and they cannot be estimated through the existing observational data in this paper. We can only explore the parameter space combined by these two parameters to determine the SFR range for each specific galaxy. In Figure \ref{Figure_7}, we show the plot of SFR vs. PA. With the SFRs from \citet{Catalan-Torrecilla-15}, 79 data points for galaxies of \citet{Yu-Ho19} are included with their error bars. The area between the black solid (upper limit) and dotted (lower limit) lines denotes the SFR range estimated by our model, and most of the data points are within this area except for several ones above the upper limit at the low PA part (The upper limit curve is the connecting line of the maximal estimating SFR of all individual galaxies, and so is the lower limit curve). This shows that the parameter space composed of $q_{2c}$ and $q_{\rho c}$ can reasonably explain the dispersion among these data points, i.e., the differences of galactic ambient conditions cause the differences of SFRs. This, in turn, shows the rationality of our model. The rising trend of SFR with increasing PA is both shown on the upper and lower limit curves, which is a rational theoretical relationship predicted by our model. However, it is difficult to be reflected on those dispersive data points. The subsequent analysis of our model can reveal the reason for this deviation between the theory and the observation. The observed pitch angle ($\theta$), shear rate ($\Gamma$), number of shocks ($n_s$), local Keplerian velocity ($V_{\rm K}$), and disk mid radial distance ($r_{\rm mid}$) are shown shown in Table \ref{Table-1}. The other theoretical model parameters, such as $k$, $q_{2c}$, and $q_{\rho c}$, are tabulated in columns (7-9) of Table \ref{Table-1}, respectively. We also mention the corresponding pre-shock ($q_{2-}$) and post-shock ($q_{2+}$) rotational velocity in column 10 and 11, respectively. We find that due to the star formation, the post-shock velocity is always lower compared to pre-shock velocity for all the cases. It implies that gas loses its angular momentum to shift towards lower angular momentum orbits and settle down after star formation (see equation \ref{ang_diss}). The maximum and minimum SFR are shown in column 12 of Table \ref{Table-1} for SFE = 100\%. We also tabulated the available observed SFR (SFR$^{\rm obs}$) for spiral galaxies from \citet{Catalan-Torrecilla-15} in column 13 of Table \ref{Table-1}.

By further analysis, more physical insights can be obtained from Figure \ref{Figure_7}. The dynamics of outflow gas from the galaxy's gaseous disk are not included in our model, so we have no way of knowing how the gas is left, but there are only two possibilities. One is the disk wind, and the other is the star formation. Unfortunately, we cannot determine or constrain the individual fractions of disk wind or SFR on the total outflow through the observational data in this paper. In fact, it is equivalent to assuming 100\% of the compressed outflow gas forms stars that the outflow rate calculated with equation (\ref{mass_out_eqn}) is directly regarded as SFR. Obviously, it is impossible to achieve a star formation efficiency (SFE) of 100\% in reality; thus, we also draw a series of upper limit curves under the different SFEs (The other colored curves paralleling the black curve) in Figure \ref{Figure_7}. This shows the differences in SFEs among these galaxies, which may be caused by the galactic ambient conditions manifested by parameters $q_{2c}$ and $q_{\rho c}$ in our model. We speculate that compared with the SFR induced by spiral arms, those galaxies close to the 1\% curve might have large disk wind launched from arms, while those beyond the 100\% curve (even including those beyond 50\%) might have other stronger star formation mechanisms. These speculations need further observations and simulations to confirm, but it is beyond the scope of our study in this paper.

\section{Discussions and Conclusions}
\label{conclusion}

In this paper, we assume a non-axisymmetric, inviscid, self-gravitating accretion flow around a compact object. We calculate the spiral shocks in the flow following \citet{Chakrabarti-90b} prescription. We also adopt the same solution methodology, i.e., a point-wise self-similar approach based on our earlier work \citep{Aktar-etal21}. In general, the matter distribution in the disk should be considered via the Poissons equation. However, it is challenging to handle analytically to get the self-gravitating effect. As a result, we consider a simplified relation between disk surface density and gravitational potential due to self-gravity \citep{Mestel-63, Lodato-07}. Our self-gravitating model immediately reduces to \citet{Chakrabarti-90b} model in the absence of self-gravity, i.e., when $\sigma =0$ (see equation \ref{potential_eqn}). First, we compare the flow variables of accretion in terms of spiral coordinates for non self-gravitating and self-gravitating disks, shown in Figure \ref{Figure_1}. We observe that the evolution of flow variables is completely different in the presence of self-gravitating disks. In the same spirit, we compare the solution topology in Figure \ref{Figure_2}. We find that the flow exhibits spiral shocks for self-gravitating disks even though there is no shock for non self-gravitating disks for the same set of flow parameters. We also observe that two-shock and four-shock solutions are possible in the presence of self-gravitating disks (see Figure \ref{Figure_2} and Figure \ref{Figure_3}). Moreover, we observe that mass outflow rates increase with the increase of pitch angle, which indicates that gaseous mass can be easily escaped from the disk due to spiral shock for weakly wound spiral arm, depicted in Figure \ref{Figure_4}. 

Further, we compare and examine the overall shock parameter space separated by pitch angle $(\theta)$ and rotational velocity at sonic surface $(q_{2c})$ by varying the number of shocks $(n_s)$. We observe that the shock parameter space shrinks with the increase of the number of shocks, shown in Figure \ref{Figure_5}. Finally, we attempt to calculate SFR for 79 spiral galaxies based on our accretion-ejection model. Interestingly, we observe that mass outflow triggered by spiral shock waves serves as one of the essential physical mechanisms for SFR, depicted in Figure \ref{Figure_7}. Moreover, our model-calculated SFR is consistent with the observed SFR for various spiral galaxies. 

Looking back at the analysis of PA-SR and SFR-PA correlation, depicted in Figure \ref{Figure_6} and Figure \ref{Figure_7}, which are both shown as very dispersive relationships by observational data. We conclude that the dispersion of data also contains rich physical information, which needs to be extracted by appropriate theoretical models, and our one is just a simple attempt in this regard. We also admit that the physical mechanism of SFR is extremely complex. There are various other mechanisms, such as AGN feedback, supernova, etc., that may trigger SFR in galaxy \citep{Salome-etal16, Padoan-etal17, Mukherjee-etal18, Cosentino-etal22}.  

In this work, we avoid any dissipation mechanisms in the disk. However, in a realistic situation, various dissipation is present in the flow. In a complete scenario, we need to incorporate the viscous effect along with the various cooling mechanisms to get the complete picture, and it will change the flow dynamics \citep{Chakrabarti-Das04, Aktar-etal17}. Moreover, it is already pointed out that the dynamics of spiral shocks are significantly affected in the presence of radiative losses in the disk \citep{Spruit-87}. Also, the radiative processes are very significant in galactic disks. The present work investigates the spiral shock properties in radiatively inefficient galactic disks (i.e., adiabatic). On the other hand, the gravitational instabilities redistribute the angular momentum in the disk \citep{Binney-Tremaine87, Bertin-Lodato99}. Further, we do not consider the gravity torque effect in our model, which is essential in spiral galactic disk \citep{Block-etal-02, Block-etal-04, Tiret-Combes-08}. Moreover, one of the major limitations of the point-wise self-similar approach is that it is impossible to investigate the global radial variations of flow variables. To analyze this more rigorously, we need a time-dependent simulation study. This kind of study is beyond our scope in the present formalism. We hope to address these issues in the future.

\section*{Acknowledgments}
 We thank the anonymous referee for very useful comments and suggestions that improved the quality of the paper. The authors also want to express their humble gratitude to Si-Yue Yu for various fruitful discussions during the preparation of the manuscript. The work was supported by the Natural Science Foundation of Fujian Province of China (No. 2023J01008).



%
%

\bibliographystyle{aa}
\bibliography{refs.bib}

\newpage
\begin{table*}
\setlength{\tabcolsep}{1.2mm}
\renewcommand\arraystretch{1.0}
\caption{Spiral galaxy and their star formation rate (SFR)
\label{Table-1}}
 \begin{tabular}{@{}c c c c c c c c c c c c c c c c } 
 \hline
 Galaxy Name &     \multicolumn{5}{c}{Observational} & &  & & & Theoretical &  &  &  &  & \\ 
      \cline{2-6}    \cline{8-13}  
   & $\theta$ & $\Gamma$ & $n_s$ & $V_{\rm K} (r_{\rm mid})$ & $r_{\rm mid}$ & & $k$ & $q_{2c}$ & $q_{\rho c}$ & $q_{2-}$ & $q_{2+}$ & ${\rm SFR}^{\rm min}$ & ${\rm SFR}^{\rm Obs}$\\ 
   &  &  &  &  &  &  &  &  &  &  & & ${\rm SFR}^{\rm max}$  \\
   & $(^\circ)$ &  & & (km/sec) & (kpc)  &  &  &  &  &  & & ($M_{\odot}/{\rm yr}$) & ($M_{\odot}/{\rm yr}$) \\   
 \hline
 IC 1151 & 23.7 & 0.74 & 3 & 135.08 & 5.10 &  & 1.7313 & 0.0800 & 0.0036 & 0.5503 & 0.1519 & 0.3606 & 0.87 $\pm$ 0.15   \\ 
        &  &  &  &  &  &  &  & 0.0650 & 0.0349 & 0.9049  & 0.1499 & 6.5473 &   \\
IC 1256 & 23.5 & 0.68 & 2 & 229.78 & 9.25 & & 1.8858 & 0.0550 & 0.0590 & 0.5934  & 0.1877 & 0.4572 & 1.47 $\pm$ 0.25   \\
        &      &      &   &        &      & &        & 0.0600 & 0.0100 & 0.9116 & 0.1431 & 6.3740    \\
IC 4566 & 16.2 & 1.04 & 2 & 218.56 & 17.47 & & 1.5833 & 0.050 & 0.0078 & 0.5135  & 0.2154 & 0.3745 & 0.47 $\pm$ 0.08   \\
        &      &      &   &        &       & &        & 0.0700 & 0.0200 & 0.8923  & 0.3431 & 2.9765    \\
IC 5309 & 14.9 & 0.99 & 2 & 140.89 & 7.70 & & 1.9167 & 0.120 & 0.0222 & 0.7870 & 0.3229 & 0.0873 & 0.78 $\pm$ 0.13   \\
        &      &      &   &        &       & &        & 0.0800 & 0.0500 & 0.7946 & 0.1580 & 2.1599    \\
MCG-02-51-004 & 14.8 & 0.59 & 2 & 270.06 & 10.72 & & 3.4442 & 0.250 & 0.030 & 0.5934 & 0.1588 & 0.0979 & 1.97 $\pm$ 0.35 \\ 
        &      &      &   &        &       & &        & 0.0300 & 0.0400 & 0.8140 & 0.1322 & 2.3249    \\
NGC 1 & 19.9  & 1.12  & 2  & 187.68  & 12.60  &   & 1.0497  & 0.1300  & 0.025 & 0.7178 & 0.3906 & 0.3194 & 3.92 $\pm$ 1.09   \\
      &       &      &   &        &       & &        & 0.1500 & 0.0700 & 0.8187 & 0.2584 & 4.2163    \\
NGC 23 & 22.8 & 1.06 & 2 & 200.76 & 20.48 & &  1.0467 & 0.1200 & 0.035 & 0.6361 & 0.2075  & 0.4737 & 12.01 $\pm$  2.04  \\ 
       &      &      &   &        &       & &        & 0.1700 & 0.0450 & 0.8307 & 0.2697 & 5.9368   \\
NGC 160 & 9.0 & 1.28 & 4 & 294.81 & 6.45 & & 1.3890 & 0.055  & 0.065  & 0.5365 & 0.1801  & 0.0083 & 0.43 $\pm$ 0.05 \\
       &      &      &   &        &       & &        & 0.1200 & 0.0600 & 0.8810 & 0.1154 & 0.4989   \\
NGC 171 & 18.2 & 0.83 & 2 & 143.82 & 13.35  &  & 2.0378 & 0.1800  & 0.025 & 0.6688 & 0.1930 & 0.3226 & 0.90 $\pm$ 0.27    \\
       &      &      &   &        &       & &        & 0.1600 & 0.0350 & 0.7411 & 0.1228 & 3.8708   \\
NGC 214 & 22.2 & 1.04 & 3 & 215.06 & 10.54 & & 1.6907 & 0.065  & 0.027  & 0.4854 & 0.2011  & 0.3194 & 3.73 $\pm$ 0.63 \\
       &      &      &   &        &       & &        & 0.1200 & 0.0350 & 0.7738 & 0.2723 & 5.8740    \\
NGC 237 & 21.2 & 0.81 & 2 & 196.22 &  9.41 & & 1.7789 & 0.1700  & 0.058  & 0.6244 & 0.2774 & 0.3363 & 2.69 $\pm$ 0.45  \\
       &      &      &   &        &       & &        & 0.0800 & 0.06500 & 0.8124& 0.3020 & 5.4942   \\
NGC 257 & 13.4 & 0.86 & 2 & 225.06 & 12.55 & & 2.6864 & 0.1400  & 0.065  & 0.4946 & 0.1509  & 0.0947 & 4.84 $\pm$ 0.82  \\
       &      &      &   &        &       & &        & 0.0800 & 0.05300 & 0.7897 & 0.1231 & 1.5831   \\ 
NGC 551 & 25.7 & 0.68 & 3 & 214.21 & 12.21 & & 1.7038 & 0.1900  & 0.0550  & 0.6768  & 0.3356  & 0.0467 & 1.80 $\pm$ 0.51 \\
       &      &      &   &        &       & &        & 0.1700 & 0.0650 & 0.7834 & 0.1003 & 7.6809    \\  
NGC 768 & 17.5 & 0.88 & 2 & 248.16 & 15.28 & & 1.9663 & 0.0560  & 0.0450  & 0.4820  & 0.2437  & 0.1650 & 3.68 $\pm$ 0.74   \\
       &      &      &   &        &       & &        & 0.0650 & 0.0370 & 0.6951 & 0.1573 & 3.2785   \\ 
NGC 776 & 20.6 & 0.91 & 3 & 143.66 & 13.38 & & 1.5696 & 0.1300  & 0.0170  & 0.5479 & 0.2633 & 0.2151 & 3.31 $\pm$ 0.55   \\
       &      &      &   &        &       & &        & 0.1150 & 0.0340 & 0.9103 & 0.1216 & 4.4963    \\ 
NGC 932 & 7.9 & 1.11 & 3 & 209.26 & 9.11 & & 2.8105 & 0.0560  & 0.0270  & 0.4626 & 0.2126 & 0.0058  & ---  \\
       &      &      &   &        &       & &        & 0.1600 & 0.0350 & 0.7895 & 0.1511 & 0.2853    \\
NGC 1167& 10.5 & 1.06 & 2 & 340.80 & 16.39 & & 2.3740 & 0.0580 & 0.0380 & 0.6364 & 0.2449 & 0.0112 & ---  \\
       &      &      &   &        &       & &        & 0.1200 & 0.0450 & 0.7404 & 0.1321 & 0.5894    \\
NGC 1349 & 5.4  & 1.01 & 2 & 305.42 & 11.56 &  & 5.1836 & 0.0560  & 0.0156  & 0.5981 & 0.2394 & 0.0005 & ---   \\
       &      &      &   &        &       & &        & 0.17000 & 0.0240 & 0.9296 & 0.4431 & 0.0535    \\
NGC 1645 & 8.6  & 1.20 & 2 & 227.79 & 15.99 &  & 1.9836 & 0.0750  & 0.0350  & 0.7342 & 0.2960 &  0.0005 & 0.65 $\pm$ 0.04  \\ 
       &      &      &   &        &       & &        & 0.0800 & 0.0240 & 0.8765 & 0.1619 & 0.42925    \\
NGC 2253 & 15.1 & 1.34 & 2 & 173.18 &  7.87 &  & 0.5929 & 0.1250  & 0.0650  & 0.5518 & 0.2171  & 0.0337 & 1.13 $\pm$ 0.19  \\
       &      &      &   &        &       & &        & 0.1600 & 0.07400 & 0.8677 & 0.1967 & 2.5401    \\ 
NGC 2449 & 21.4 & 0.88 & 2 & 227.10 & 11.19 &  & 1.5820 & 0.1050 & 0.0500 & 0.8750  & 0.3318  & 0.0372 &  0.75 $\pm$ 0.14  \\
         &      &      &   &        &       &  &        & 0.2200 & 0.0720 & 0.7211  & 0.2062  & 5.1998 &  \\ 
NGC 2486 & 11.0 & 1.04 & 2 & 191.87 & 10.38 &  & 2.3664 & 0.0650 & 0.0160 & 0.7143  & 0.2528  & 0.0006 & ---   \\
         &      &      &   &        &       &  &        & 0.1450 & 0.0405 & 0.8042  & 0.2658  & 0.9954 \\ 
NGC 2604 & 24.3 & 0.63 & 3 & 134.08 &  7.87 &  & 1.9268 & 0.0350 & 0.0125 & 0.6699  & 0.1925  & 0.0059 &  --- \\
         &      &      &   &        &       &  &        & 0.1070 & 0.0205 & 0.7449  & 0.1433  & 7.5877 \\ 
NGC 2730 & 20.1 & 0.52 & 2 &  78.53 &  1.97 &  & 2.6779 & 0.0560 & 0.0650 & 0.7943 & 0.4211   & 0.0032 &  1.62 $\pm$ 0.27\\
         &      &      &   &        &       &  &        & 0.1450 & 0.0570 & 0.9055 & 0.2676   & 4.5054 &\\ 
NGC 2906 & 28.2 & 0.75 & 2 & 233.61 &  6.17 &  & 1.3987 & 0.0450 & 0.0620 & 0.6792 & 0.2885   & 0.0066 & 0.68 $\pm$ 0.11\\
         &      &      &   &        &       &  &        & 0.1250 & 0.1050 & 0.8453 & 0.2262   & 8.9611 &\\ 
NGC 2916 & 19.2 & 0.85 & 2 & 242.24 & 10.18 &  & 1.8665 & 0.0750 & 0.0235 & 0.7309 & 0.3154   & 0.0027 & 1.90 $\pm$ 0.33\\
         &      &      &   &        &       &  &        & 0.1250 & 0.0350 & 0.8248 & 0.2450   & 3.9493 &\\ 
NGC 3057 & 20.1 & 0.69 & 2 &  59.55 &  1.50 &  & 2.2134 & 0.0850 & 0.0270 & 0.5961 & 0.2914   & 0.0030 & 0.22 $\pm$ 0.03\\
         &      &      &   &        &       &  &        & 0.2150 & 0.0150 & 0.8897 & 0.2727   & 4.2734 &\\ 
NGC 3106 & 12.2 & 1.19 & 2 & 211.64 & 16.08 &  & 1.4338 & 0.0645 & 0.0280 & 0.7309 & 0.3193   & 0.0008 & 0.79 $\pm$ 0.10\\
         &      &      &   &        &       &  &        & 0.1550 & 0.0350 & 0.8815 & 0.2382   & 1.1668 &\\ 
NGC 3381 & 26.6 & 0.86 & 2 &  84.22 &  5.06 &  & 1.2780 & 0.0450 & 0.0360 & 0.6993 & 0.2871   & 0.0059 & 0.41 $\pm$ 0.07\\
&      &      &   &        &       &  &        & 0.1750 & 0.1050 & 0.9045 & 0.2706   & 8.0294 &\\
 \hline
 \end{tabular}
\end{table*}


\begin{table*}
\setlength{\tabcolsep}{1.2mm}
\renewcommand\arraystretch{1.0}
 \begin{tabular}{@{}c c c c c c c c c c c c c c c c } 
  \multicolumn{14}{l}{{\bf Table 1} -- {\it continued}} \\
 \hline
 Galaxy Name &     \multicolumn{5}{c}{Observational} & &  & & & Theoretical &  &  &  &  & \\ 
      \cline{2-6}    \cline{8-13}  
   & $\theta$ & $\Gamma$ & $n_s$ & $V_K (r_{\rm mid})$ & $r_{\rm mid}$ & & $k$ & $q_{2c}$ & $q_{\rho c}$ & $q_{2-}$ & $q_{2+}$ & ${\rm SFR}^{\rm min}$ & ${\rm SFR}^{\rm Obs}$\\ 
   &  &  &  &  &  &  &  &  &  &  & & ${\rm SFR}^{\rm max}$  \\
   & $(^\circ)$ &  & & (km/sec) & (kpc)  &  &  &  &  &  & & ($M_{\odot}/{\rm yr}$) & ($M_{\odot}/{\rm yr}$) \\   
 \hline
NGC 3815 &  9.4 & 0.96 & 2 & 201.85 &  9.09 &  & 3.2618 & 0.0650 & 0.0550 & 0.6753 & 0.3117   & 0.0004 & 1.08 $\pm$ 0.18\\
         &      &      &   &        &       &  &        & 0.1650 & 0.0850 & 0.8941 & 0.2804   & 0.6037 &\\ 
NGC 3994 & 9.9 & 0.97  & 1 & 285.16 &  6.10 &  & 3.0367 & 0.0750 & 0.0105 & 0.6929 & 0.2874   & 0.0004 & 4.38 $\pm$ 1.07\\
         &      &      &   &        &       &  &        & 0.1350 & 0.0405 & 0.8257 & 0.2312   & 0.6417 &\\ 
NGC 4003 & 9.2 & 1.06  & 2 & 207.04 & 15.53 &  & 2.7166 & 0.0207 & 0.0505 & 0.7354 & 0.2440   & 0.0004 & 0.36 $\pm$ 0.03\\
         &      &      &   &        &       &  &        & 0.1050 & 0.0800 & 0.8992 & 0.2519   & 0.5613 &\\ 
NGC 4047 & 14.6& 0.93  & 2 & 252.38 &  5.11 &  & 2.1882 & 0.0305 & 0.0250 & 0.5425 & 0.2692   & 0.0016 & 3.48 $\pm$ 1.00\\
         &      &      &   &        &       &  &        & 0.1250 & 0.0650 & 0.7249 & 0.2674   & 2.3245 &\\ 
NGC 4185 & 10.9 & 0.70 & 2 & 220.23 & 13.10 &  & 4.1543 & 0.0450 & 0.0750 & 0.6701 & 0.2899   & 0.0004 & 0.85 $\pm$ 0.29\\
         &      &      &   &        &       &  &        & 0.1350 & 0.0850 & 0.7982 & 0.2578   & 0.6027 &\\ 
NGC 4210 & 27.4 & 0.56 & 3 & 208.18 &  7.27 &  & 1.8134 & 0.0450 & 0.0205 & 0.7259 & 0.2224   & 0.0063 & 0.74 $\pm$ 0.22\\
         &      &      &   &        &       &  &        & 0.0620 & 0.0840 & 0.8104 & 0.2375   & 8.6202 &\\ 
NGC 4644 & 32.5 & 0.95 & 2 & 200.56 & 10.70 &  & 0.8633 & 0.0120 & 0.0140 & 0.7259 & 0.3172   & 0.0070 & 0.52 $\pm$ 0.12\\
         &      &      &   &        &       &  &        & 0.0850 & 0.0755 & 0.8841 & 0.2535   & 10.0554 &\\ 
NGC 4711 & 25.3 & 0.60 & 4 & 169.51 &  6.92 &  & 1.9039 & 0.0230 & 0.0120 & 0.7861 & 0.2810   & 0.0056 & 1.07 $\pm$ 0.27\\
         &      &      &   &        &       &  &        & 0.0950 & 0.0850 & 0.9006 & 0.3449   & 7.7504 &\\ 
NGC 4961 & 36.5 & 0.55 & 3 & 186.58 &  5.26 &  & 1.2838 & 0.0750 & 0.0340 & 0.6933 & 0.2912   & 0.0079 & 0.58 $\pm$ 0.15\\
         &      &      &   &        &       &  &        & 0.1650 & 0.0670 & 0.8104 & 0.2562   & 11.6224 &\\ 
NGC 5000 & 23.3 & 0.99 & 2 & 131.56 & 16.42 &  & 1.3931 & 0.0340 & 0.0210 & 0.6396 & 0.3009   & 0.0051 & 1.57 $\pm$ 0.41\\
         &      &      &   &        &       &  &        & 0.0850 & 0.0950 & 0.8771 & 0.2589   & 7.0359 &\\ 
NGC 5056 & 26.3 & 0.90 & 3 & 208.74 & 13.62 &  & 1.2140 & 0.0450 & 0.0230 & 0.7122 & 0.2856   & 0.0061 &  2.89 $\pm$ 0.48\\
         &      &      &   &        &       &  &        & 0.1350 & 0.0650 & 0.7982 & 0.2298   & 8.3085 &\\
NGC 5614 & 9.1  & 1.08 & 1 & 268.84 &  8.90 &  & 2.6221 & 0.0630 & 0.0125 & 0.6982 & 0.2869   & 0.0012 & 0.64 $\pm$ 0.11 \\
         &      &      &   &        &       &  &        & 0.1020 & 0.0750 & 0.7836 & 0.2389   & 1.2535 &  \\
NGC 5633 & 23.4 & 0.64 & 3 & 260.73 &  6.51 &  & 1.9873 & 0.0450 & 0.0235 & 0.6986 & 0.3212   & 0.0037 & 1.78 $\pm$ 0.48\\
         &      &      &   &        &       &  &        & 0.0765 & 0.0355 & 0.8784 & 0.2672   &  14.5416 \\
NGC 5657 & 19.6 & 1.10 & 2 & 191.96 &  3.01 &  & 1.1233 & 0.0345 & 0.0125 & 0.6166 & 0.1936   & 0.0040 & 1.77 $\pm$ 0.57  \\
         &      &      &   &        &       &  &        & 0.1160 & 0.0765 & 0.7513 & 0.1603   & 8.1092  \\
NGC 5720 & 7.5 & 0.86  & 2 & 272.55 & 18.84 &  & 4.8612 & 0.0550 & 0.0235 & 0.5681 & 0.3173   & 0.0029 & 0.96 $\pm$ 0.20 \\
         &      &      &   &        &       &  &        & 0.1250 & 0.0865 & 0.8841 & 0.2395   & 1.0555 & \\
NGC 5732 & 14.4 & 0.80 & 2 & 175.32 &  7.06 &  & 2.7263 & 0.0755 & 0.0234 & 0.6743 & 0.2030   & 0.0051 & 0.94 $\pm$ 0.23   \\
         &      &      &   &        &       &  &        & 0.1230 & 0.0565 & 0.7774 & 0.1885   & 4.7776 & \\
NGC 5876 & 9.9 & 1.35  & 2 & 198.14 & 12.30 &  & 0.8594 & 0.0234 & 0.0375 & 0.6693 & 0.2158   & 0.0049 & 0.15 $\pm$ 0.02   \\
         &      &      &   &        &       &  &        & 0.0850 & 0.0735 & 0.7955 & 0.1825   & 1.3573 &    \\
NGC 5888 & 10.2 & 0.78 & 2 & 292.90 & 14.37 &  & 4.0015 & 0.0475 & 0.0235 & 0.6728 & 0.2902   & 0.0052 & 1.24 $\pm$ 0.25   \\
         &      &      &   &        &       &  &        & 0.1350 & 0.0785 & 0.8471 & 0.2705   & 1.2080 & \\
NGC 5890 & 19.5 & 0.82 & 2 & 242.52 &  8.42 &  & 1.9202 & 0.0354 & 0.0315 & 0.6658 & 0.3017   & 0.0086 &   ---             \\
         &      &      &   &        &       &  &        & 0.0965 & 0.0675 & 0.7569 & 0.2671   & 7.7592 & \\
NGC 6060 & 22.5 & 0.71 & 2 & 239.28 & 13.67 &  & 1.9072 & 0.0475 & 0.0165 & 0.6930 & 0.2101   & 0.0149 &  6.73 $\pm$ 1.14  \\
         &      &      &   &        &       &  &        & 0.0895 & 0.0745 & 0.7873 & 0.1226   & 12.1004 &\\
NGC 6063 & 14.1 & 0.54 & 2 & 160.60 &  5.52 &  & 3.8219 & 0.0750 & 0.0475 & 0.6282 & 0.2948   & 0.0051 &  0.49 $\pm$ 0.08  \\
         &      &      &   &        &       &  &        & 0.1540 & 0.0865 & 0.8011 & 0.1247   & 4.2776 & \\
NGC 6186 & 21.3 & 0.92 & 2 & 136.04 &  4.07 &  & 1.4876 & 0.0326 & 0.0176 & 0.7097 & 0.3147   & 0.0149 &  1.50 $\pm$ 0.44  \\
         &      &      &   &        &       &  &        & 0.0854 & 0.0732 & 0.8487 & 0.1804   & 11.1004&   \\
NGC 6301 & 15.5 & 0.55 & 4 & 258.82 & 18.15 &  & 3.4255 & 0.0356 & 0.0324 & 0.6148 & 0.2595   & 0.0179 &  4.81 $\pm$ 0.87  \\
         &      &      &   &        &       &  &        & 0.0954 & 0.1045 & 0.9008 & 0.3094   & 5.5102 &   \\
NGC 6478 & 18.2 & 0.89 & 3 & 284.46 & 18.38 &  & 1.8553 & 0.0135 & 0.0243 & 0.7125 & 0.2811   & 0.0194 &  3.34 $\pm$ 1.11  \\
         &      &      &   &        &       &  &        & 0.0745 & 0.0650 & 0.8311 & 0.2765   & 8.6973 &   \\
NGC 6497 & 12.0 & 1.12 & 2 & 240.33 & 12.65 &  & 1.7877 & 0.0345 & 0.0254 & 0.7125 & 0.2823   & 0.0040 &  0.39 $\pm$ 0.07  \\
         &      &      &   &        &       &  &        & 0.0785 & 0.0456 & 0.8419 & 0.1673   & 2.7292 & \\
NGC 6941 & 13.1 & 1.12 & 2 & 207.71 & 16.44 &  & 1.6329 & 0.0175 & 0.0354 & 0.6012 & 0.2024   & 0.0042 &  0.33 $\pm$ 0.06  \\
         &      &      &   &        &       &  &        & 0.0765 & 0.0856 & 0.7913 & 0.2495   & 3.4464 & \\
NGC 7047 & 30.1 & 0.57 & 2 & 241.79 & 10.73 &  & 1.6043 & 0.0325 & 0.0175 & 0.6544 & 0.2412   & 0.0013 &  2.04 $\pm$ 0.34  \\
         &      &      &   &        &       &  &        & 0.0956 & 0.0743 & 0.7795 & 0.1127   & 19.7103&   \\
NGC 7311 & 9.0  & 1.12 & 2 & 251.18 & 11.48 &  & 2.3992 & 0.0215 & 0.0450 & 0.6270 & 0.2164   & 0.0002 &  2.06 $\pm$ 0.56  \\
         &      &      &   &        &       &  &        & 0.0765 & 0.0564 & 0.9001 & 0.1774   & 1.2619 & \\
NGC 7321 & 14.3 & 0.95 & 3 & 270.35 & 17.45 &  & 2.1577 & 0.0325 & 0.0275 & 0.6770 & 0.3218   & 0.0058 &  3.51 $\pm$ 0.59  \\
         &      &      &   &        &       &  &        & 0.0850 & 0.0735 & 0.8062 & 0.1522   & 5.0587 &\\
 \hline
 \end{tabular}
\end{table*}


\begin{table*}
\setlength{\tabcolsep}{1.2mm}
\renewcommand\arraystretch{1.0}
 \begin{tabular}{@{}c c c c c c c c c c c c c c c c } 
  \multicolumn{14}{l}{{\bf Table 1} -- {\it continued}} \\
 \hline
 Galaxy Name &     \multicolumn{5}{c}{Observational} & &  & & & Theoretical &  &  &  &  & \\ 
      \cline{2-6}    \cline{8-13}  
   & $\theta$ & $\Gamma$ & $n_s$ & $V_K (r_{\rm mid})$ & $r_{\rm mid}$ & & $k$ & $q_{2c}$ & $q_{\rho c}$ & $q_{2-}$ & $q_{2+}$ & ${\rm SFR}^{\rm min}$ & ${\rm SFR}^{\rm Obs}$\\ 
   &  &  &  &  &  &  &  &  &  &  & & ${\rm SFR}^{\rm max}$  \\
   & $(^\circ)$ &  & & (km/sec) & (kpc)  &  &  &  &  &  & & ($M_{\odot}/{\rm yr}$) & ($M_{\odot}/{\rm yr}$) \\   
 \hline
 NGC 7364 & 12.7 & 1.16 & 2 & 269.43 &  7.25 &  & 1.5086 & 0.0435 & 0.0275 & 0.5907 & 0.2269   & 0.0049 &  3.00 $\pm$ 0.51  \\
         &      &      &   &        &       &  &        & 0.0925 & 0.0825 & 0.8243 & 0.2429   & 3.6042 & \\
NGC 7466 & 26.3 & 0.97 & 3 & 216.24 & 16.95 &  & 1.0723 & 0.0215 & 0.0354 & 0.6970 & 0.2316   & 0.0214 & 2.85 $\pm$ 0.49   \\
         &      &      &   &        &       &  &        & 0.0785 & 0.0567 & 0.7604 & 0.2172   & 18.6428&\\
NGC 7489 & 29.4 & 0.72 & 4 & 202.99 & 16.21 &  & 1.3842 & 0.0565 & 0.0445 & 0.6747 & 0.2397   & 0.0022 & 10.22 $\pm$ 1.80  \\
         &      &      &   &        &       &  &        & 0.1350 & 0.0856 & 0.7532 & 0.1843   & 19.0381& \\
NGC 7591 & 16.7 & 1.11 & 2 & 188.45 & 13.23 &  & 1.2999 & 0.0135 & 0.0295 & 0.5869 & 0.2122   & 0.0016 & 13.68 $\pm$ 2.53  \\
         &      &      &   &        &       &  &        & 0.2450 & 0.0845 & 0.7496 & 0.2700   & 6.0845 &\\
NGC 7631 & 25.1 & 1.00 & 3 & 186.71 &  9.55 &  & 1.0673 & 0.0456 & 0.0235 & 0.6487 & 0.2503   & 0.0020 & 0.93 $\pm$ 0.15   \\
         &      &      &   &        &       &  &        & 0.1750 & 0.1050 & 0.8517 & 0.1397   & 15.5281& \\
NGC 7653 & 29.0 & 1.02 & 3 & 200.01 & 10.05 &  & 0.8659 & 0.0565 & 0.0275 & 0.6471 & 0.1827   & 0.0021 & 2.59 $\pm$ 0.43   \\
         &      &      &   &        &       &  &        & 0.1650 & 0.0856 & 0.8291 & 0.1724   & 18.6784&\\
NGC 7716 & 27.3 & 1.11 & 4 & 141.09 &  7.98 &  & 0.7556 & 0.0475 & 0.0345 & 0.5241 & 0.2367   & 0.0020 & 0.65 $\pm$ 0.11   \\
         &      &      &   &        &       &  &        & 0.0955 & 0.0565 & 0.8803 & 0.1743   & 17.4610&\\
NGC 7738 & 11.0 & 1.02 & 2 & 200.13 & 28.71 &  & 2.4693 & 0.0345 & 0.0105 & 0.6368 & 0.2367   & 0.0042 & 8.49 $\pm$ 1.44   \\
         &      &      &   &        &       &  &        & 0.0756 & 0.0805 & 0.7556 & 0.1615   & 1.3464 & \\
NGC 7819 & 30.8 & 0.87 & 2 & 139.91 & 12.55 &  & 1.0568 & 0.0575 & 0.0206 & 0.7418 & 0.2569   & 0.0024 & 2.07 $\pm$ 0.35   \\
         &      &      &   &        &       &  &        & 0.1250 & 0.0785 & 0.8754 & 0.1746   & 20.0524& \\
NGC 7824 & 6.3  & 1.20 & 2 & 277.84 & 15.42 &  & 2.7173 & 0.0850 & 0.0355 & 0.5562 & 0.2085   & 0.0001 &  ---           \\
         &      &      &   &        &       &  &        & 0.1750 & 0.1050 & 0.8803 & 0.3280   & 0.1606 & \\
UGC 5    & 19.7 & 0.67 & 2 & 242.48 & 10.77 &  & 2.4577 & 0.0455 & 0.0235 & 0.5052 & 0.1481   & 0.0061 & 4.10 $\pm$ 0.71   \\
         &      &      &   &        &       &  &        & 0.0854 & 0.0765 & 0.7049 & 0.1643   & 8.0112 & \\
UGC 1271 & 14.2 & 1.20 & 2 & 244.27 &  4.13 &  & 1.1855 & 0.0256 & 0.0155 & 0.5581 & 0.2790   & 0.0046 &  ---           \\
         &      &      &   &        &       &  &        & 0.0875 & 0.0655 & 0.7410 & 0.1987   & 4.3752 & \\
UGC 2403 & 14.2 & 0.74 & 2 & 217.27 & 14.36 &  & 3.0034 & 0.0125 & 0.0345 & 0.5778 & 0.2102   & 0.0055 & 2.70 $\pm$ 0.45   \\
         &      &      &   &        &       &  &        & 0.0745 & 0.0675 & 0.6984 & 0.1861   & 4.4908 & \\
UGC 3253 &  4.2 & 0.87 & 1 & 211.50 &  9.01 &  & 8.5789 & 0.0235 & 0.0102 & 0.6611 & 0.2443   & 0.0001 & 0.96 $\pm$ 0.27   \\
         &      &      &   &        &       &  &        & 0.0956 & 0.0785 & 0.7672 & 0.2378   & 0.1016 & \\
UGC 4398 & 25.2 & 0.65 & 3 & 159.64 &  9.77 &  & 1.8063 & 0.0245 & 0.0405 & 0.7015 & 0.2067   & 0.0017 &  ---           \\
         &      &      &   &        &       &  &        & 0.1250 & 0.1065 & 0.7570 & 0.1987   & 15.4750& \\
UGC 7012 & 20.0 & 0.81 & 2 & 107.80 &  0.84 &  & 1.8957 & 0.0256 & 0.0345 & 0.6265 & 0.2067   & 0.0014 &  0.65 $\pm$ 0.17  \\
         &      &      &   &        &       &  &        & 0.1150 & 0.0956 & 0.7111 & 0.1680   & 8.5146 &   \\
UGC 7145 & 21.3 & 0.72 & 2 & 188.39 & 10.36 &  & 2.0005 & 0.0375 & 0.0157 & 0.6960 & 0.2324   & 0.0015 &   ---           \\
         &      &      &   &        &       &  &        & 0.1350 & 0.1150 & 0.8155 & 0.2272   & 13.9998& \\
UGC 8781 & 10.7 & 1.10 & 2 & 222.76 & 18.65 &  & 2.1169 & 0.0246 & 0.0345 & 0.6295 & 0.2102   & 0.0014 &  1.18 $\pm$ 0.25  \\
         &      &      &   &        &       &  &        & 0.0785 & 0.1750 & 0.7672 & 0.1943   & 1.4002 & \\
UGC 9476 & 31.7 & 0.64 & 2 & 151.97 &  6.91 &  & 1.3924 & 0.0565 & 0.0345 & 0.6108 & 0.2492   & 0.0023 &  1.28 $\pm$ 0.35   \\
         &      &      &   &        &       &  &        & 0.1350 & 0.0765 & 0.7797 & 0.2256   & 20.8836&   \\
UGC 10796& 30.8 & 0.81 & 2 & 118.52 & 11.67 &  & 1.1574 & 0.0750 & 0.0235 & 0.7062 & 0.2351   & 0.0022 &  0.21 $\pm$ 0.04   \\
         &      &      &   &        &       &  &        & 0.2150 & 0.0956 & 0.8081 & 0.2832   & 19.6163& \\
UGC 12185& 15.5 & 1.12 & 2 & 215.01 & 14.81 &  & 1.3702 & 0.0345 & 0.0215 & 0.5711 & 0.2199   & 0.0060 &  0.88 $\pm$ 0.17   \\
         &      &      &   &        &       &  &        & 0.0855 & 0.0955 & 0.8318 & 0.2565   & 5.5402  & \\
 \hline
 \end{tabular}
\end{table*}

\end{document}